\documentclass[sigconf]{acmart}
\usepackage{booktabs} 
\usepackage{multirow}
\usepackage{graphicx}
\usepackage{balance}
\usepackage{subfigure}
\usepackage{bm}
\usepackage{amsmath}
\usepackage{amsfonts}
\usepackage{graphicx}

\usepackage[ruled,vlined]{algorithm2e}
\usepackage{algpseudocode}

\newcommand{\squishlist}{
 \begin{list}{$\bullet$}
  { \setlength{\itemsep}{0pt}
     \setlength{\parsep}{3pt}
     \setlength{\topsep}{3pt}
     \setlength{\partopsep}{0pt}
     \setlength{\leftmargin}{1.5em}
     \setlength{\labelwidth}{1em}
     \setlength{\labelsep}{0.5em} } }

\newcommand{\squishlisttwo}{
 \begin{list}{$\bullet$}
  { \setlength{\itemsep}{0pt}
     \setlength{\parsep}{0pt}
    \setlength{\topsep}{0pt}
    \setlength{\partopsep}{0pt}
\setlength{\leftmargin}{2em}
\setlength{\labelwidth}{1.5em}
\setlength{\labelsep}{0.5em} } }

\newcommand{\squishend}{
\end{list}  }

\copyrightyear{2022} 
\acmYear{2022} 
\setcopyright{rightsretained} 
\acmConference[WWW '22]{Proceedings of the ACM Web Conference 2022}{April 25--29, 2022}{Virtual Event, Lyon, France}
\acmBooktitle{Proceedings of the ACM Web Conference 2022 (WWW '22), April 25--29, 2022, Virtual Event, Lyon, France}\acmDOI{10.1145/3485447.3512147}
\acmISBN{978-1-4503-9096-5/22/04}

\begin{document}
\title{Learning to Augment for Casual User Recommendation}

\author{Jianling Wang*}\thanks{*Work was done as an intern at Google.}
\affiliation{%
  \institution{Texas A\&M University}
  \country{College Station, TX, USA}
}
\email{jlwang@tamu.edu}

\author{Ya Le}
\affiliation{%
  \institution{Google, Inc.}
  \country{Mountain View, CA, USA}
}
\email{elainele@google.com}

\author{Bo Chang}
\affiliation{%
  \institution{Google, Inc.}
  \country{Mountain View, CA, USA}
}
\email{bochang@google.com}

\author{Yuyan Wang}
\affiliation{%
  \institution{Google, Inc.}
  \country{Mountain View, CA, USA}
}
\email{yuyanw@google.com}

\author{Ed H. Chi}
\affiliation{%
  \institution{Google, Inc.}
  \country{Mountain View, CA, USA}
}
\email{edchi@google.com}

\author{Minmin Chen}
\affiliation{%
  \institution{Google, Inc.}
  \country{Mountain View, CA, USA}
}
\email{minminc@google.com}



\renewcommand{\shortauthors}{Wang, Jianling, et al.}
\begin{abstract}
Users who come to recommendation platforms are heterogeneous in activity levels. There usually exists a group of core users who visit the platform regularly and consume a large body of content upon each visit, while others are casual users who tend to visit the platform occasionally and consume less each time.
As a result, consumption activities from core users often dominate the training data used for learning. As core users can exhibit different activity patterns from casual users, recommender systems trained on historical user activity data usually achieve much worse performance on casual users than core users. 
To bridge the gap, we propose a model-agnostic framework \textit{L2Aug} to improve recommendations for casual users through data augmentation, without sacrificing core user experience. \textit{L2Aug} is powered by a data augmentor that learns to generate augmented interaction sequences, in order to fine-tune and optimize the performance of the recommendation system for casual users. On four real-world public datasets, \textit{L2Aug} outperforms other treatment methods and achieves the best sequential recommendation performance for both casual and core users. We also test \textit{L2Aug} in an online simulation environment with real-time feedback to further validate its efficacy, and showcase its flexibility in supporting different augmentation actions.

\end{abstract}
%

\ccsdesc[500]{Information systems}
\ccsdesc[500]{Information systems~Recommender systems}

\keywords{Recommendation Systems, Data Augmentation, Policy Learning}

\maketitle

\section{Introduction}

Recommendation systems are ubiquitous. For example, streaming services rely on recommender systems to serve high-quality informational and entertaining content to their users, and e-commerce platforms recommend interesting items to assist customers in making shopping decisions. Sequential recommendation, which focuses on predicting the next item the user is interested in consuming based on their historical interaction sequences, has been extensively studied in many applications \cite{hidasi2015session,kang2018self,tang2018personalized,yuan2019simple,hidasi2018recurrent}. 

\begin{figure*}[!t]
\vspace{-0.1in}
    \graphicspath{{figures/}}
    \centering
    \scalebox{0.99}{
    \subfigure[]
    {
    \hspace{-0.1cm}
    \includegraphics[width=0.33\textwidth]{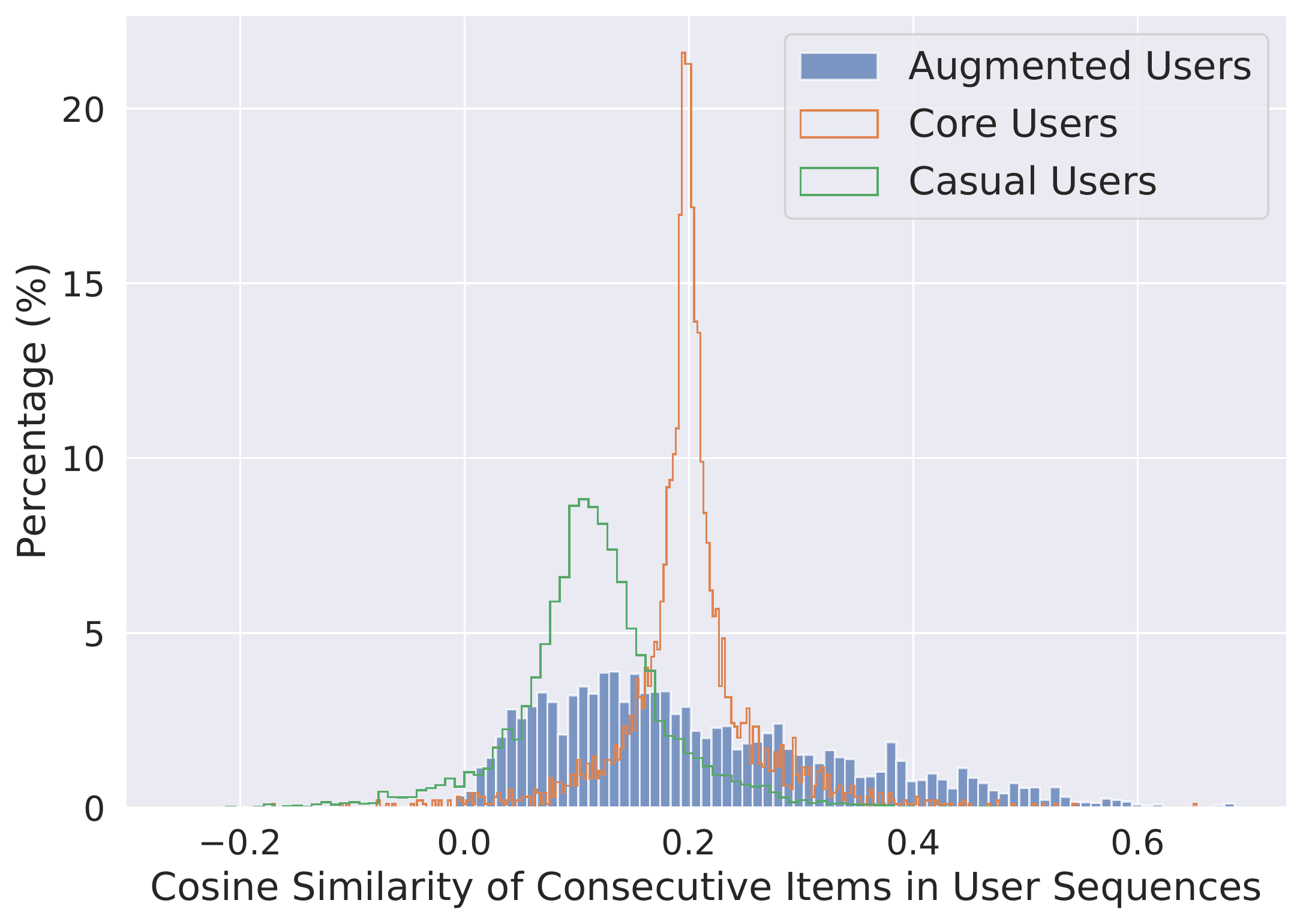}
    }
    \hspace{-0.1cm}
    \subfigure[]
    {
    \includegraphics[width=0.335\textwidth]{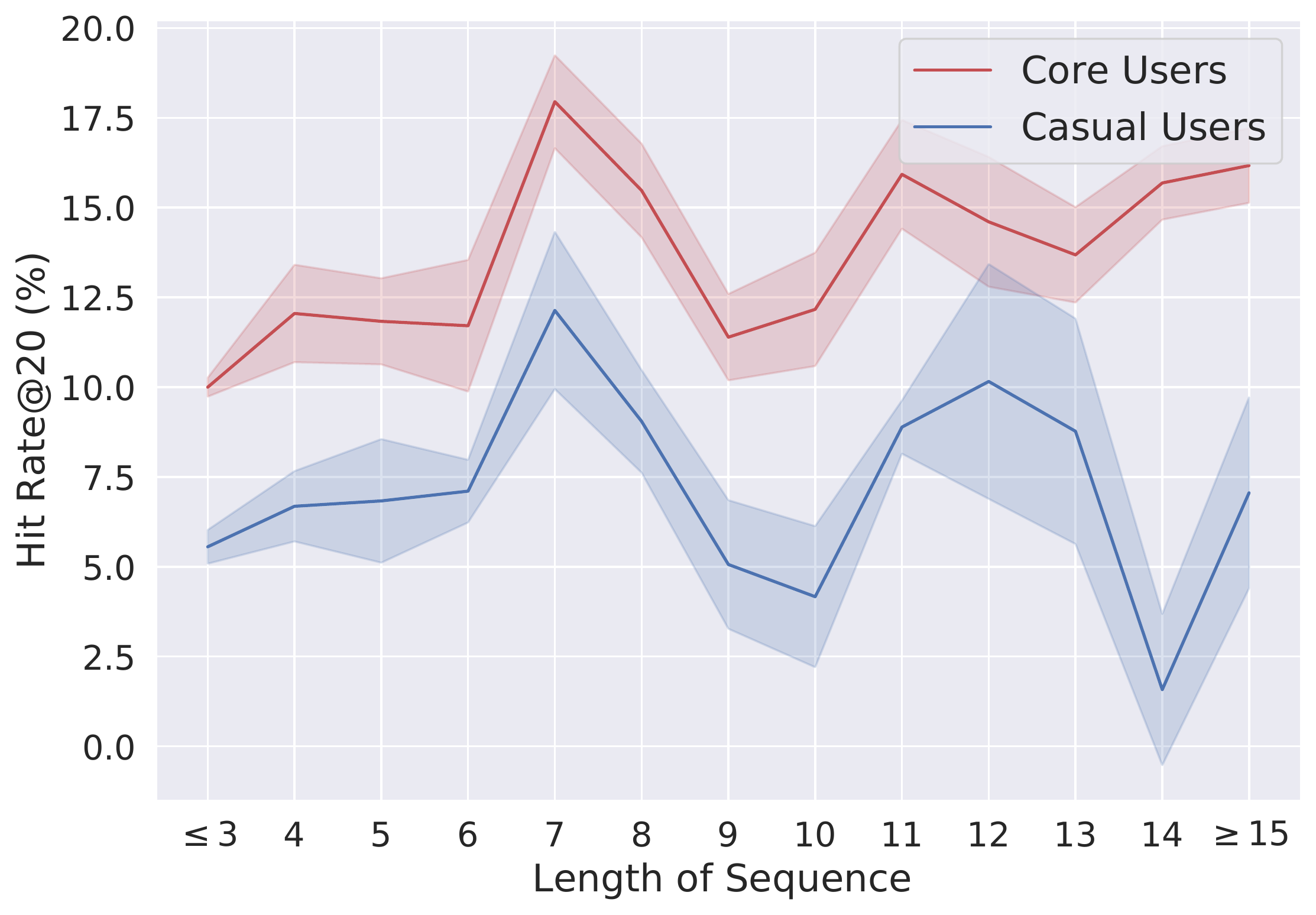}
    }
    \hspace{-0.1cm}
    \subfigure[]
    {
    \includegraphics[width=0.325\textwidth]{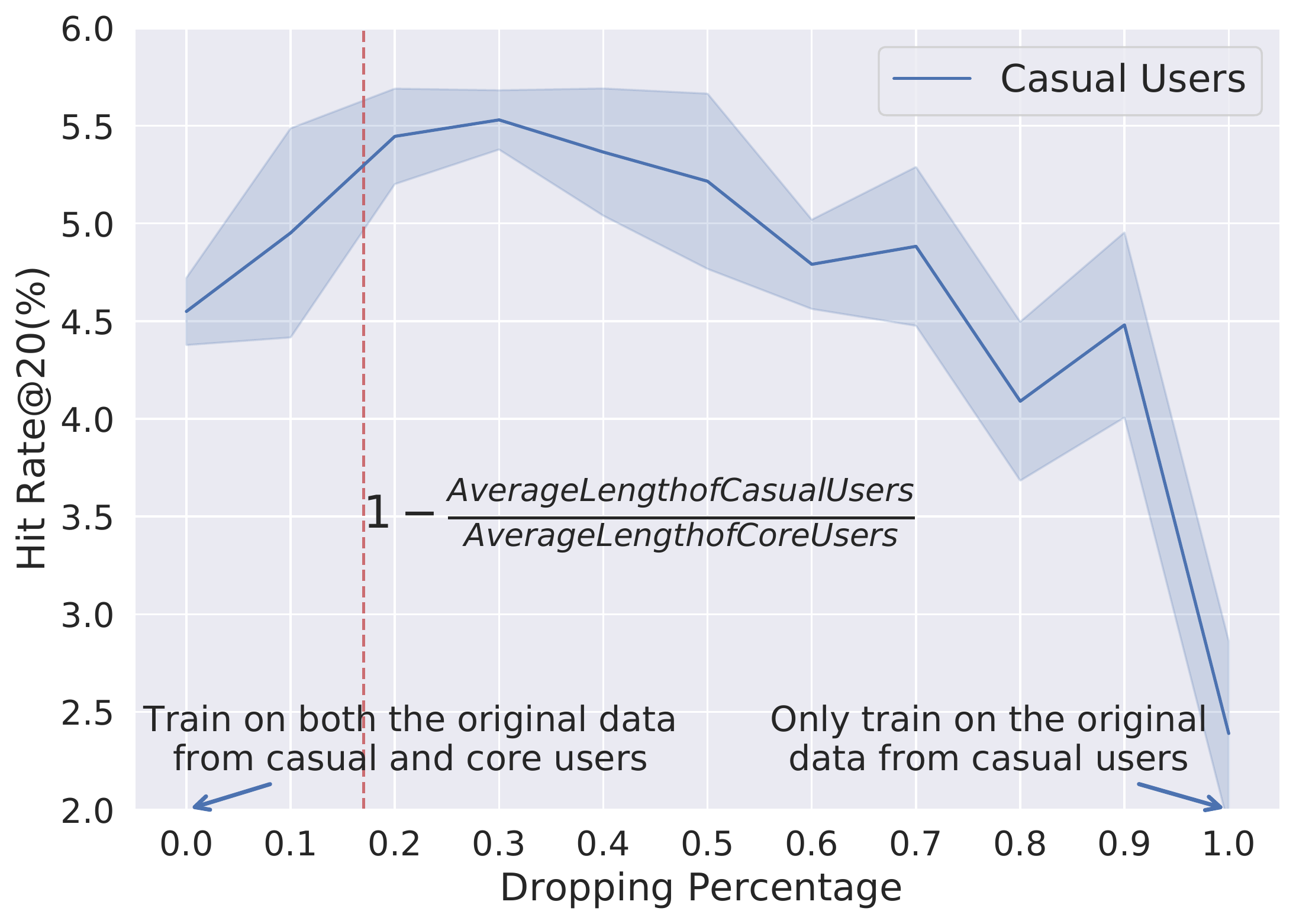}
    }}
    \vspace{-0.05in}
    \caption{(a) Comparison of interest continuity (i.e., the similarity of items consumed consecutively) for different user groups. (b) Given a recommendation system (SASRec \cite{kang2018self}) trained on interaction sequences from both core and casual users, for interaction sequences of the same length, it always performs worse on casual users than on core users, indicating the difficulty of making recommendations for casual users. (c) Data from core users is helpful for training a model for casual users. Furthermore, randomly dropping part of the interactions from core users can further improve the performance on casual users.}%
    \vspace{-0.05in}
    \label{fig:preliminary}
    \label{fig:explain}
    \label{fig:viz}
\end{figure*}

Users coming to the online platform are often heterogeneous in activity levels. There usually exists a set of \textbf{\textit{core users}} who visit the platform regularly and consistently, while others are \textbf{\textit{casual users}} who tend to visit the platform occasionally. The heterogeneity in activity levels can lead to distinct transitional patterns between these two groups of users \cite{buttle2004customer,chen2021values}. As shown in Figure \ref{fig:viz}(a), consecutively interacted items are less concentrated, and of lower similarity in casual users than core as they come to the platform less frequently. 

Sequential recommenders trained predominantly on interaction data from core users often fail to capture the activity patterns of casual users and, as a result, provide less satisfactory recommendations for casual users. 
As shown in Figure \ref{fig:explain}(b), the self-attention based recommender (SASRec \cite{kang2018self}) performs significantly worse on casual users than on core users in all sequence lengths. 
\textit{How to improve the recommendation for casual users without sacrificing the performance on core users} is a critical challenge for building satisfactory recommendation services for all.

Existing methods such as cold-start recommendation \cite{gantner2010learning,li2019zero,volkovs2017dropoutnet,zheng2020cold} or cross-domain approaches \cite{zhu2021cross,khan2017cross,ganin2016domain,zhao2020catn,kim2017learning}, mainly focus on addressing the data scarcity, but fail to handle the activity disparity between different types of users in the system. Although there are often more casual users than core users on the platforms, they left much fewer interactions in total compared to the core users. How to distill informative transition patterns from core users and efficiently adapt to casual users is the main research question we aim to tackle here. 
Inspired by the recent advances in data augmentation techniques \cite{cubuk2019autoaugment,krizhevsky2012imagenet,ko2015audio,wei2019eda}, we set out to generate augmented data sequences from core users to mimic the behavior patterns of casual users. While many augmentation techniques have been studied for continuous inputs such as images~\cite{chen2020simple,he2020momentum}, augmentation on user activity data consisting of sequential discrete item IDs is still under-explored. Meanwhile, compared with core users that tend to leave consistent and informative interaction sequences, casual users usually have noisier and more diverse behavior sequences as shown in Figure~\ref{fig:preliminary}(a). It is an open question to find an effective data augmentation method to generate augmented sequences that inherit informative transition patterns from core users and improve casual user recommendations.



To tackle the aforementioned challenges, we propose a model-agnostic ``Learning to Augment'' framework -- \textit{L2Aug}, which bridges the gap between casual and core users for sequential recommendation systems. Specifically, we develop a data augmentor which decides on a series of augmentation actions on the input sequence to generate augmented data sequences. From the ``Learning to Augment'' perspective, this data augmentor is trained to conduct effective data augmentation to maximize the performance of the \textit{target model} (i.e., recommender). 
Framing this as learning a data augmentation policy, the data augmentor (agent) generates context/state and chooses the augmentation action.
Meanwhile, the target model is updated with the augmented data sequences, and its performance improvement on a meta validation set is used as the reward to guide the learning of the augmentation policy.
Through alternating between the data augmentation step using the recommender performance as the reward, and improving the recommender with the augmented data, the two modules reinforce each other and progressively improve both the data augmentation and recommendation quality.
As a result, this builds an adaptive recommendation system, which can distill informative transition patterns from core users and adapt to casual users with dramatically different interaction patterns. Our contributions can be summarized as follows:
\squishlist
\item We investigate the disparity between core and casual users in their transitional patterns within the interaction sequences, and study the feasibility of bridging the gap between sequential recommendation for core and casual users from the data augmentation perspective.
\item We propose a model-agnostic framework \textit{L2Aug} to learn a data augmentation policy using REINFORCE and improve the recommendation system using generated augmented data.
\item We evaluate \textit{L2Aug}, on top of various SOTA sequential recommendation models, on four real-world datasets, and show that it outperforms other treatment methods and achieves the best recommendation performance on both core and casual users. 
\item We also evaluate \textit{L2Aug} in an online simulation environment, where the user responses on counterfactual recommendations are known. We also showcase its effectiveness as well as flexibility in supporting multiple augmentation actions.
\squishend

\section{Motivation}
\label{sec:motivation}
In this section, we conduct an initial investigation with data sampled from the public Amazon review dataset \cite{mcauley2015image}, to explore the distinct behavior patterns between casual and core users, and then examine the feasibility of applying data augmentation in bridging the gap between them. 

Firstly, to investigate the interest continuity in item consumption history of different users, we compute the \textit{correlation between consecutive items} in their interaction sequences. We apply the bag-of-words model on item descriptions to obtain the item embeddings and then calculate the cosine similarity between the embeddings of consecutive items in the interaction sequences. In Figure \ref{fig:viz}(a), we can observe that the consecutive items consumed by core users are more similar. It confirms our hypothesis that core and casual users behave differently and the interests of casual users are less concentrated compared with core users.

Next, as interaction sequences of core users tend to be longer and denser than those of casual users, the most straightforward approach for data augmentation is to \textit{randomly drop} part of the interactions from core users. We adopt this approach in this initial investigation and train a SASRec model \cite{kang2018self} with the augmented data.
In Figure \ref{fig:preliminary}(c), we visualize the performance for casual user recommendation by varying the percentage of dropped interactions. When the dropping percentage is equal to 0, none of the interactions from core users is dropped, thus the recommender is trained on the original data from both core and casual users. On the contrary, when the dropping percentage is equal to 1.0, all interactions from the core users are dropped, meaning that the recommender is trained only on the original data from casual users. It can be observed that the recommender system achieves improved performance on casual users when we start to drop interactions from the core users, which suggests that the synthetic data can help improve casual user recommendation. 
However, as the dropping percentage increases, discarding too much information negatively impacts casual user recommendation. These observations motivate us to search for more fine-grained and controlled augmentation policies. 

\begin{figure*}
\vspace{-0.1in}
\includegraphics[width=0.97\textwidth]{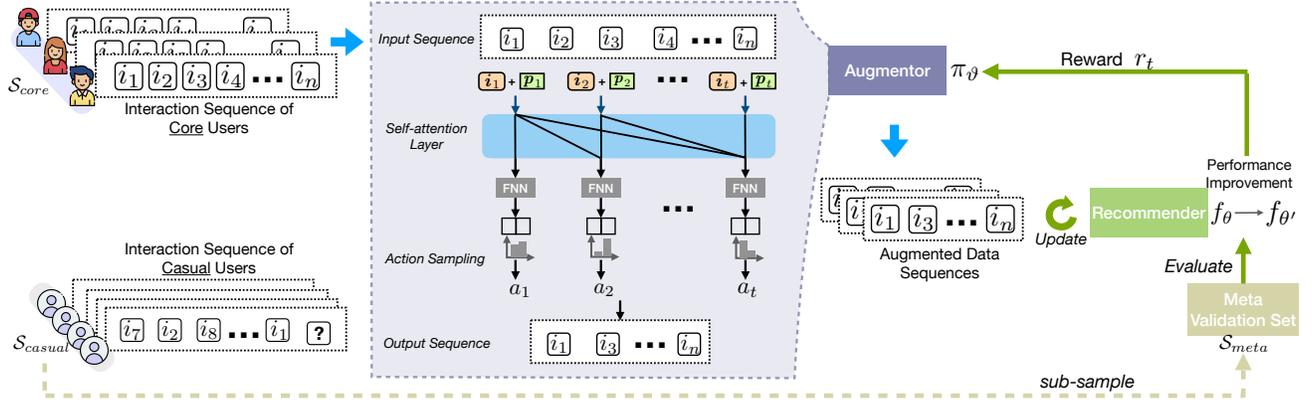}
\caption{The proposed model-agnostic ``Learning to Augment'' framework -- \textit{L2Aug}. The \textit{data augmentor} takes a series of augmentation \underline{\textit{actions}} on the input sequence to generate synthetic data sequences. Meanwhile the \textit{recommender} is updated by the synthetic data sequences; its performance improvement on a meta validation set is used as the \underline{\textit{reward}} to guide the training of the augmentor. These two components reinforce each other and progressively improve both the data augmentation and recommendation quality. }
\vspace{-0.05in}
\label{fig:model}
\end{figure*}

\section{Preliminaries}
In this section, we describe the problem setting of sequential recommendation and introduce the baseline recommendation model.

\smallskip
\noindent\textbf{Problem Formulation.}
We use $\mathcal{U}=\{u_1, u_2, ..., u_{|\mathcal{U}|}\}$ and $\mathcal{I}=\{i_1, i_2, ..., i_{|\mathcal{I}|}\}$ to denote the set of $|\mathcal{U}|$ users and the set of $|\mathcal{I}|$ items on the platform. In this work, the users on the platform can be partitioned into two groups: casual users $\mathcal{U}_{casual}$ and core users $\mathcal{U}_{core}$  based on their activity levels (i.e., the frequency they visit the platform), s.t. $\mathcal{U}_{casual} \cap \mathcal{U}_{core} = \emptyset$ and $\mathcal{U}_{casual} \cup \mathcal{U}_{core} = \mathcal{U}$. Note that each item is mapped to a trainable embedding vector associated with its unique ID. In this work, we do not consider any auxiliary information for users and items. Given the sequence of items that user $u$ has interacted with in chronological order $\textbf{S}_u = [i_{u,1}, i_{u,2},\ldots, i_{u,p}, \ldots, i_{u,n}]$, where $i_{u,p}$ represents the $p$th item $u$ interacted with,
the objective of a sequential recommendation model is to infer the next interesting item $i_{u,{n+1}}$ for user $u$. 
To simplify notations, we use $\textbf{S}_{u,[1:p]}$ to denote the subsequence of items $[i_{u,1}, i_{u,2},\ldots, i_{u,p}]$.

\smallskip
\noindent\textbf{Target Model -- Recommender.}
We learn a target model (recommender) $f$, which predicts the preference score for each candidate item in $\mathcal{I}$ given $\textbf{S}_{u,[1:j-1]}$:
\begin{equation}
    \begin{aligned}
        \hat{\mathbf{y}}_{u,j} = f(\textbf{S}_{u,[1:j-1]};\theta),
    \end{aligned}
    \label{equ:inference}
\end{equation}
where $\theta$ represents the parameters of the function $f$ and the prediction $\hat{\mathbf{y}}_{u,j} \in \mathcal{R}^{|\mathcal{I}|}$ denotes the predicted scores of user $u$ at step $j$ on all the candidates in $\mathcal{I}$. Then the items with the top preference scores based on $ \hat{\textbf{y}}_{u,j}$ are recommended to user $u$. Specifically, we adopt the cross-entropy loss as the objective function, that is:
\begin{equation}
    \begin{aligned}
        \ell(\theta) = -\sum_{u\in\mathcal{U}}
        \sum_{j=1}^{|\textbf{S}_u|}\left[\log(\sigma(\hat{\textbf{y}}_{u,j}^{i_{u,j}}))+\sum_{k\notin\textbf{S}_u}\log(1-\sigma(\hat{\textbf{y}}_{u,j}^k))\right],
    \end{aligned}
\end{equation}
where $\sigma$ is the Sigmoid function and $\hat{\textbf{y}}_{u,j}^{i_{u,j}}$ is the predicted preference score for the ground-truth item $i_{u,j}$ that the user $u$ interacted with at step $j$. In our experiments, we sample a single $k\notin\textbf{S}_u$ as the negative sample for each user at each step $j$.

\section{Learning to Augment}
In this section, we elaborate on the design of \textit{L2Aug}, organized around two guiding research questions: 1) How to perform data augmentation on sequential data to support various operations (e.g., remove, keep or substitute) on items within the sequence? 2) How to learn an effective data augmentation policy to achieve the goal of improving causal user recommendation?

\smallskip
\noindent\textbf{Framework Overview.} We propose \textit{L2Aug}, illustrated in Figure \ref{fig:model}, to learn the data augmentation policy for improving casual user recommendation. There are two main components in the design: a \textit{recommender} to make sequential recommendations and a \textit{data augmentor} to generate synthetic interaction sequences by applying the learned data augmentation policy on the input sequences.
These two components are alternately trained. 
Specifically, for each batch of sequences sampled from $\textbf{S}_{core}$, the \textit{data augmentor} learns to take a series of augmentation actions (e.g., remove, keep or substitute) so the generated synthetic sequences can improve the performance of the \textit{recommender}. 
Framing this as learning a data augmentation policy, the data augmentor (agent) generates context/state and chooses the augmentation action.
Meanwhile, the target model is updated with the augmented data sequences and its performance improvement on a meta validation set is used as the reward to guide the training of the augmentor.
Through alternating between the data augmentation step using the recommender performance as the reward, and improving the recommender with the augmented data, the two modules reinforce each other and progressively improve both the data augmentation and recommendation quality.

\subsection{Learning Augmentation Policy}

Our goal is to \textit{learn a discrete sequential data augmentation policy} for \textit{maximizing the performance of recommendation systems} on casual users. Inspired by preliminary studies in Section~\ref{sec:motivation}, we set out to generate augmented (synthetic) data sequences to mimic the behavior patterns of casual users by editing the core user sequences. For each batch of interaction sequences sampled from $\mathcal{U}_{core}$, we consider as a \textit{data augmentation task} that generates the synthetic sequences by taking a series of editing actions from \{keep, drop, replace, ...\} for items in the input sequences sequentially. Since the data augmentation process is non-diffentiable, we adopt the policy learning framework described below to enable the training:

\squishlist

\item \textbf{Context/State}: Let $\textbf{S}_u$ denote an interaction sequence from $\mathcal{U}_{core}$. When encountering item $i_{u,k}$ in sequence $\textbf{S}_u$, a vector $\textbf{h}_k$ that encodes the subsequence $\textbf{S}_{[1:k]}$ is regarded as the context or state representation. The detailed model to obtain the state vector will be discussed in Section \ref{sec:da}. 

\item \textbf{Action}: For simplicity, we use two actions -- ``Keep'' and ``Drop''. 
The model could be extended to support multiple actions (i.e., more than two); more details can be found in Section \ref{sec:more_actions}. At step $k$, for item $i_{u, k}$ in sequence $\textbf{S}_u$, we need to decide on the action $a_{u,k} \in \{0,1\}$ to keep or drop item $i_{u,k}$ when generating the augmented sequence from sequence $\textbf{S}_u$. Note that $a_{u,k} = 0$ indicates dropping the item and $a_{u,k} = 1$ means to keep it. 

\item \textbf{Meta Valication Set ($\mathcal{U}_{meta}$)}: To guide the training of the data augmentor, from the set of $N$ casual users $\mathcal{U}_{casual}$, we randomly sample a small subset of $M$ users ($M \ll N$) and construct the meta validation set $\mathcal{U}_{meta}$ with their interaction sequences. During the training process, the performance of the target model on $\mathcal{U}_{meta}$ is computed as the reward for learning the augmentation policy.

\item \textbf{Reward}: The reward function aims to guide the agent to learn the augmentation policy in order to maximize the performance of the target model. Each batch of the augmented sequences is used to train the target model and leads to a performance change of the target model, which can be regarded as the reward to the data augmentor. In offline settings, it can be the \textit{update of recommendation performance} on $\mathcal{U}_{meta}$ measured by offline metrics (e.g., NDCG, Hit rate and Mean Reciprocal Ranking). In online settings where we can simulate user response on counterfactual recommendations, the reward can be the user response returned by the environment (e.g., engagement, rating or conversion). More details about the metrics can be found in Section~\ref{sec:experimental_setup}. In our experiments, we use the performance gain on both NDCG and Hit Rate as the reward for offline experiment, and use the change of simulated rating as the reward for online experiment. Following~\cite{yao2019coacor, zhang2020selective,ding2021learning}, the policy network for data augmentation is updated on a delayed reward received after feeding the generated augmented data to the recommender.

\squishend

\subsection{Data Augmentor}
\label{sec:da}
Since most of the existing data augmentation methods are designed for continuous feature spaces~\cite{chen2020simple,he2020momentum}, they are not fit for handling sequence data consisting of discrete item IDs in our case. We propose the \textit{Data Augmentor}, which generates a synthetic sequence by encoding the input sequence as the \textbf{context/state} representations, and deciding on the editing \textbf{actions} on the input sequence. 

Given a sequence of interacted items $\textbf{S}_u = [i_{u,1}, \ldots, i_{u,k}, \ldots, i_{u,t}]$ for a core user $u \in \mathcal{U}_{core}$, the Data Augmentor needs to decide on the editing action on each item. To make the decision on each item $i_{u, k}$, the agent needs to encode the subsequence $\text{S}_{[1:k]}$, which in turn requires a representation of each item. We encode two pieces of information in the individual item representation: the item content itself and its position. In other words, for each item $i_{u, k}$ in $\textbf{S}_u$, we have $\textbf{e}_{k} = \textbf{i}_{k} + \textbf{p}_k$. Here, $\textbf{i}_{k}$ is the embedding for item $i_{u,k}$ and $\textbf{p}_k$ is the positional embedding of position $k$, which is used to retain the order information.

With the individual item representation, any sequential embedding model including RNN, Bidirectional-RNN or Transformers \cite{vaswani2017attention,hochreiter1997long} can be used to encode the subsequence  $\text{S}_{[1:k]}$ to produce a context/state representation.
In this work, we adopt the self-attention model \cite{vaswani2017attention} to produce the state $\textbf{h}_k$.  Self-attention \cite{vaswani2017attention} is designed to match a sequence against itself and thus uses the same objects as the queries, keys, and values. We will transform the position-aware embeddings with $\textbf{W}_Q$, $\textbf{W}_K$, and $\textbf{W}_V$ to generate the query vectors, key vectors, and value vectors, correspondingly. Then we can generate the state representation for each item $i_{u, k}$ by encoding its correlation with other items in the sequence via:
\begin{equation}
\textbf{h}_k = \sum_{j \leq k} \sigma_{kj} \textbf{W}_V\textbf{e}_{j}
\quad\mbox{and}\quad
\sigma_{kj} = \frac{S(\textbf{W}_Q\textbf{e}_{k}, \textbf{W}_K\textbf{e}_{j})}{\sum_{j \leq k} S(\textbf{W}_Q\textbf{e}_{k}, \textbf{W}_K\textbf{e}_{j})},
\end{equation}
in which the function $S(\cdot,\cdot)$ is used to denote the similarity score between two elements. In particular, we use the \textit{Scaled Dot-Product Attention} \cite{kang2018self,vaswani2017attention} to calculate the scores:
$S(\textbf{a}, \textbf{b}) = {\textbf{a}^T\textbf{b}}/{\sqrt{D}},$
where $D$ is the dimension size and used for normalization when calculating the similarity scores. Once obtaining each individual \textbf{\textit{context/state}} representation $\textbf{h}_{k}$, the agent produces a policy $\pi_k$ over the action space $\mathcal{A}$ through
\begin{equation}
\pi_{\bm\vartheta}(\cdot|\mathbf{h}_k) = \text{Softmax}(\textbf{W}_A\cdot\textbf{h}_k). 
\end{equation}
Here $\textbf{W}_A \in \mathcal{R}^{|\mathcal{A}|\times |\textbf{h}|}$ is a trainable weight matrix. We use $\bm\vartheta$ to denote all the parameters involved in building the augmentation policy network. Each dimension in $\pi_{\bm\vartheta,k}$ represents the probability of a specific action at step $k$. The agent then decides on the editing \textbf{\textit{action}} for the item by sampling with the probabilities.


\begin{algorithm}[t]
\caption{\textit{L2Aug} Training Process}
\label{alg:L2Aug}
\LinesNumbered
\small
\KwIn{Training sequences $\mathcal{S}_{core}$, $\mathcal{S}_{meta}$,  $\mathcal{S}_{casual}$ and the pre-trained recommendation system $f_{\bm\theta}$, update frequency $\digamma$}
\KwOut{A fine-tuned recommendation system $f_{\bm\theta}$}
Random initialize the data augmentor $\pi_{\bm\vartheta}$ and $i=0$.

\While{not converge}{

    Sample $B$ samples $\mathcal{S}_B$ from $\mathcal{S}_{core}$

    // \texttt{Augmentation by Data Augmentor}
    
    \For{$u = 1 \rightarrow B$} {
        \For{$k = 1 \rightarrow |\textbf{s}_u|$} {
        Compute the state representation for the $k_{th}$ item
        
        Sample the augmentation action
        
        }
        
        Obtain the augmented sequence
    }

    // \texttt{Augmentation Policy Optimization}
    
    Fine-tune $f_{\bm\theta}$ with the batch of augmented sequences to get $f_{\bm\theta'}$
    
    Calculate the reward $r_t = \text{HT}(f_{\bm\theta'},\mathcal{S}_{meta}) - \text{HT}(f_{\bm\theta},\mathcal{S}_{meta})$
    
    Update $\pi_{\bm\vartheta}$ according to Eq. (\ref{eq:update}) and \eqref{eq:gradient}
    
    // \texttt{Replay and Update the Recommender}
    
    \If{$i \bmod \digamma == 0$}{
    Sample $B$ samples $\mathcal{S}_{train}$ from $\mathcal{S}_{core}$
    
    Generate synthetic samples $\mathcal{D}_{syn}$ from $\mathcal{S}_{train}$ with $\pi_{\bm\vartheta}$
    
    Sample $B$ samples $\mathcal{S}_{train}'$ from $\mathcal{S}_{casual}$ and $\mathcal{S}_{core}$
    
    Update the recommender $f_{\bm\theta}$ with $S_{syn}$ and  $\mathcal{S}_{train}'$
    
    }
    
    $i \gets i + 1$
    }

    \Return The fine-tuned recommendation system $f_{\bm\theta}$

\end{algorithm}

\subsection{Augmentation Policy Optimization} To optimize for the data augmentation policy, we aim to maximize expected rewards on improved recommendation quality, which can be defined as:
\begin{equation}
    J(\bm\vartheta) = \max_{\bm\vartheta}\ \mathbb{E}_{\pi_{\bm\vartheta}}\left[ r_t\right],
\end{equation}
where $r_t$ is the reward computed as the improvement in recommendation performance when feeding the augmented data generated from input batch $t$. In Algorithm \ref{alg:L2Aug} line 12, we use the Hit Rate \text{HT} to evaluate the recommendation performance and the difference of the two as the reward. Note that Hit Rate can be replaced with any performance metrics or their combination. We update the parameters $\bm\vartheta$ via policy gradient:
\begin{equation}
    \label{eq:update}
    \bm\vartheta \leftarrow \bm\vartheta + \alpha \nabla_{\bm\vartheta} \Tilde{J}(\bm\vartheta),
\end{equation}
in which $\alpha$ is the learning rate. With the obtained rewards, the gradient can be computed by:
\begin{equation}
\label{eq:gradient}
    \nabla_{\bm\vartheta}  \Tilde{J}(\bm\vartheta) =  r_t \sum_{u \in \mathcal{S}_B}\sum_{k=1}^{|\textbf{S}_u|}\nabla_{\bm\vartheta} \log \pi_{\bm\vartheta} (a_{u,k}|\mathbf{h}_k),
\end{equation}
Details of the training process are shown in Algorithm \ref{alg:L2Aug}. Note that the same meta reward over the batch is assigned to all the augmentation decisions taken within the batch.
In essence, the obtained reward based on recommendation improvement computed on the meta validation set is used to guide the learning of the data augmentation policy. The augmented sequences in return are used to further improve the performance of the recommender.  
During the training process, the data augmentor and recommender system can reinforce each other, and progressively improve the data augmentation and recommendation quality.

\smallskip
\noindent\textbf{Replay.} To ensure that the model still achieves satisfying performance on core users, we adopt the replay strategy \cite{rolnick2018experience} to avoid forgetting. Besides the synthetic sequences, the recommendation system is also updated with the original data sequences from core users, leading to a recommender that has improved performance on casual users without sacrificing the performance on core users. 

\section{Experiments}

In this section, we report our experiments over multiple datasets to evaluate the performance of \textit{L2Aug} and answer the following questions: (i) Whether \textit{L2Aug} improves recommendations for casual users without sacrificing the performance on core users? (ii) Is the proposed framework flexible to support more augmentation actions and various recommendation setups?

\subsection{Experimental Setup}
\label{sec:experimental_setup}

\smallskip
\noindent\textbf{Datasets.} To examine the performance of the proposed method, we conduct experiments on four real-world datasets. Table~\ref{tab:data} shows the summary statistics for each dataset. \textbf{\textit{Amazon\_CDs}} is adopted from the public Amazon review dataset \cite{mcauley2015image}, which includes reviews spanning May 1996 -- July 2014 on products belonging to the ``CDs and Vinyl'' category on Amazon. Similarly, \textbf{\textit{Amazon\_Books}} and \textbf{\textit{Amazon\_Movies}} are from two of the largest categories -- ``Books'' and ``Movies'' of the same Amazon review dataset. To further investigate the performance in other application scenarios, we include another public dataset from an idea-sharing community -- \textbf{\textit{Goodreads}}, on which users can leave their reviews and ratings on books \cite{wan2018item}. For all the datasets, all the items the user has \emph{interacted} with (reviewed) form the user interaction sequence $\textbf{S}_u$. We use the average time gap between their consecutive interactions to differentiate between casual and core users: core users are those with average time gaps of less than 30 days; others are labeled as casual users. More details on data prepossessing and splitting can be found in Appendices \ref{sec:preprocess}.

\begin{table}[t]
\centering
\caption{Summary statistics for the datasets.}
\scalebox{0.99}{
\begin{tabular}{ccccc}
\toprule
 & \# Items  & \begin{tabular}[c]{@{}c@{}}\# Casual \\ Users\end{tabular} &\begin{tabular}[c]{@{}c@{}}\# Core \\ Users\end{tabular} & \begin{tabular}[c]{@{}c@{}} Avg. \# \\ Interactions\end{tabular} \\ 
\midrule
\textbf{\textit{Amazon\_CDs}} & 22,685 &  2,176 & 1,022 & 23.75\\ 
\textbf{\textit{Amazon\_Books}} & 17,443  & 11,083 & 3,457 & 31.71 \\ 
\textbf{\textit{Amazon\_Movies}} & 11,079  & 10,020 & 2,808 &  14.06\\ 
\textbf{\textit{Goodreads}} & 65,864 &  11,836 &  5,017  & 130.03\\ \bottomrule
\end{tabular}%
}
\label{tab:data}
\vspace{-0.15in}
\end{table}

\begin{table*}[t!]
\vspace{-0.1in}
\centering
\caption{Performance on \underline{casual} user recommendation of various models on different datasets.}

\setlength{\tabcolsep}{2.5pt}
\scalebox{0.985}{
\begin{tabular}{@{}l cc c cc c cc c cc @{}}
\toprule

\multirow{2}{*}{\textbf{Method}} & \multicolumn{2}{c}{\bf{Amazon\_CDs}}  & &  \multicolumn{2}{c}{\bf{Amazon\_Books}}  &  & \multicolumn{2}{c}{\bf{Amazon\_Movies}}  &  & \multicolumn{2}{c}{\bf{Goodreads}} 

\\ \cline{2-3} \cline{5-6} \cline{8-9} \cline{11-12}


\rule{0pt}{10pt} & \multicolumn{1}{c}{NDCG@5 (\%)}  & \multicolumn{1}{c}{HT@5 (\%)} &&   \multicolumn{1}{c}{NDCG@5 (\%)}  & \multicolumn{1}{c}{HT@5 (\%)} & &
\multicolumn{1}{c}{NDCG@5 (\%)}  & \multicolumn{1}{c}{HT@5 (\%)} & &
\multicolumn{1}{c}{NDCG@5 (\%)}  & \multicolumn{1}{c}{HT@5 (\%)}

\\ \midrule

GRU    & $0.64\pm0.04$  & $1.05\pm0.09$  & & $0.66\pm0.04$ & $1.10\pm0.04$  &&  $1.11\pm0.07$ & $1.81\pm0.08$ && $1.32\pm0.06$ & $2.09\pm0.08$   \\

\it{w/} Random & $0.66\pm0.06$  & $1.13\pm0.11$  & & $0.71\pm0.03$ & $1.15\pm0.05$  &&  $1.22\pm0.01$ & $1.95\pm0.12$ && $1.44\pm0.13$ & $2.29\pm0.14$   \\

\it{w/} Focused & $0.69\pm0.03$  & $1.17\pm0.08$  & & $0.74\pm0.02$ & $1.19\pm0.04$  &&  $1.28\pm0.06$ & $2.08\pm0.07$ && $1.43\pm0.05$ & $2.37\pm0.11$   \\

\it{w/} Adversarial & $0.73\pm0.02$  & $1.24\pm0.06$  & & $0.73\pm0.03$ & $1.20\pm0.05$  &&  $1.34\pm0.05$ & $2.23\pm0.06$ && $1.46\pm0.04$ & $2.21\pm0.08$   \\

\it{w/} L2Aug & $\mathbf{0.80\pm0.03}$  & $\mathbf{1.37\pm0.07}$  & & $\mathbf{0.75\pm0.02}$ & $\mathbf{1.25\pm0.03}$  &&  $\mathbf{1.43\pm0.04}$ & $\mathbf{2.35\pm0.05}$ && $\mathbf{1.53\pm0.03}$ & $\mathbf{2.45\pm0.05}$   \\
\midrule

NextItNet    & $1.12\pm0.16$  & $1.69\pm0.18$  & & $0.97\pm0.03$ & $1.56\pm0.06$  &&  $1.30\pm0.05$ & $2.13\pm0.09$ && $2.05\pm0.11$ & $2.97\pm0.13$   \\

\it{w/} Random & $1.17\pm0.18$  & $1.88\pm0.20$  & & $1.08\pm0.05$ & $1.71\pm0.07$  &&  $1.45\pm0.08$ & $2.26\pm0.09$ && $2.13\pm0.10$ & $3.11\pm0.16$   \\

\it{w/} Focused & $1.22\pm0.15$  & $2.07\pm0.16$  & & $1.13\pm0.04$ & $1.68\pm0.03$  &&  $1.47\pm0.06$ & $2.42\pm0.08$ && $2.25\pm0.09$ & $3.32\pm0.13$   \\

\it{w/} Adversarial & $1.52\pm0.11$  & $2.39\pm0.13$  & & $1.15\pm0.04$ & $1.75\pm0.05$  &&  $1.58\pm0.06$ & $2.56\pm0.08$ && $2.42\pm0.06$ & $3.43\pm0.11$   \\

\it{w/} L2Aug & $\mathbf{1.62\pm0.09}$  & $\mathbf{2.44\pm0.10}$  & & $\mathbf{1.24\pm0.05}$ & $\mathbf{1.87\pm0.04}$  &&  $\mathbf{1.71\pm0.05}$ & $\mathbf{2.74\pm0.07}$ && $\mathbf{2.51\pm0.07}$ & $\mathbf{3.62\pm0.10}$   \\
\midrule

SASRec    & $1.83\pm0.10$  & $2.77\pm0.16$  & & $1.13\pm0.05$ & $1.78\pm0.06$  &&  $1.72\pm0.05$ & $2.71\pm0.08$ && $2.29\pm0.07$ & $3.49\pm0.10$   \\

\it{w/} Random & $1.85\pm0.15$  & $2.81\pm0.18$  & & $1.21\pm0.05$ & $1.86\pm0.07$  &&  $1.76\pm0.11$ & $2.73\pm0.12$ && $2.36\pm0.11$ & $3.54\pm0.19$   \\

\it{w/} Focused & $1.88\pm0.14$  & $2.90\pm0.13$  & & $1.24\pm0.03$ & $1.94\pm0.05$  &&  $1.81\pm0.09$ & $2.83\pm0.11$ && $2.42\pm0.10$ & $3.60\pm0.14$   \\

\it{w/} Adversarial & $1.92\pm0.13$  & $3.03\pm0.14$  & & $1.23\pm0.02$ & $1.93\pm0.03$  &&  $1.88\pm0.06$ & $2.86\pm0.10$ && $2.45\pm0.08$ & $3.72\pm0.11$   \\

\it{w/} L2Aug & $\mathbf{2.11\pm0.11}$  & $\mathbf{3.26\pm0.13}$  & & $\mathbf{1.31\pm0.04}$ & $\mathbf{1.99\pm0.04}$  &&  $\mathbf{1.95\pm0.05}$ & $\mathbf{3.00\pm0.08}$ && $\mathbf{2.71\pm0.09}$ & $\mathbf{3.93\pm0.10}$   \\

\bottomrule
\end{tabular}}

\medskip
\setlength{\tabcolsep}{2.5pt}
\scalebox{0.965}{
\begin{tabular}{@{}l cc c cc c cc c cc @{}}
\toprule

\multirow{2}{*}{\textbf{Method}} & \multicolumn{2}{c}{\bf{Amazon\_CDs}}  & &  \multicolumn{2}{c}{\bf{Amazon\_Books}}  &  & \multicolumn{2}{c}{\bf{Amazon\_Movies}}  &  & \multicolumn{2}{c}{\bf{Goodreads}} 

\\ \cline{2-3} \cline{5-6} \cline{8-9} \cline{11-12}


\rule{0pt}{10pt} & \multicolumn{1}{c}{NDCG@10 (\%)}  & \multicolumn{1}{c}{HT@10 (\%)} &&   \multicolumn{1}{c}{NDCG@10 (\%)}  & \multicolumn{1}{c}{HT@10 (\%)} & &
\multicolumn{1}{c}{NDCG@10 (\%)}  & \multicolumn{1}{c}{HT@10 (\%)} & &
\multicolumn{1}{c}{NDCG@10 (\%)}  & \multicolumn{1}{c}{HT@10 (\%)}

\\ \midrule

GRU    & $0.84\pm0.07$  & $1.74\pm0.13$  & & $0.92\pm0.03$ & $1.87\pm0.05$  &&  $1.56\pm0.06$ & $3.22\pm0.11$ && $1.81\pm0.07$ & $3.61\pm0.12$   \\

\it{w/} Random & $0.91\pm0.12$  & $1.92\pm0.16$  & & $0.99\pm0.04$ & $2.02\pm0.03$  &&  $1.67\pm0.08$ & $3.34\pm0.15$ && $1.94\pm0.10$ & $3.84\pm0.20$   \\

\it{w/} Focused & $0.93\pm0.06$  & $1.91\pm0.14$  & & $1.03\pm0.03$ & $2.06\pm0.05$  &&  $1.77\pm0.06$ & $3.59\pm0.13$ && $1.91\pm0.09$ & $3.89\pm0.15$   \\

\it{w/} Adversarial & $0.96\pm0.04$  & $1.97\pm0.13$  & & $0.99\pm0.04$ & $2.02\pm0.04$  &&  $1.75\pm0.08$ & $3.51\pm0.14$ && $1.93\pm0.08$ & $3.68\pm0.13$   \\

\it{w/} L2Aug & $\mathbf{1.03\pm0.03}$  & $\mathbf{2.11\pm0.10}$  & & $\mathbf{1.05\pm0.04}$ & $\mathbf{2.16\pm0.03}$  &&  $\mathbf{1.86\pm0.05}$ & $\mathbf{3.65\pm0.10}$ && $\mathbf{1.99\pm0.07}$ & $\mathbf{3.96\pm0.11}$   \\
\midrule

NextItNet    & $1.59\pm0.17$  & $2.87\pm0.20$  & & $1.29\pm0.05$ & $2.55\pm0.08$  &&  $1.79\pm0.07$ & $3.66\pm0.16$ && $2.52\pm0.12$ & $4.44\pm0.18$   \\

\it{w/} Random & $1.57\pm0.18$  & $3.10\pm0.24$  & & $1.34\pm0.07$ & $2.53\pm0.09$  &&  $2.05\pm0.13$ & $4.10\pm0.18$ && $2.66\pm0.16$ & $4.74\pm0.19$   \\

\it{w/} Focused & $1.45\pm0.12$  & $2.80\pm0.19$  & & $1.47\pm0.05$ & $2.74\pm0.06$  &&  $1.92\pm0.11$ & $3.80\pm0.14$ && $2.69\pm0.15$ & $4.73\pm0.17$   \\

\it{w/} Adversarial & $1.85\pm0.11$  & $3.44\pm0.15$  & & $1.48\pm0.04$ & $2.77\pm0.05$  &&  $2.11\pm0.09$ & $4.20\pm0.11$ && $2.89\pm0.07$ & $4.90\pm0.13$   \\

\it{w/} L2Aug & $\mathbf{2.11\pm0.12}$  & $\mathbf{3.95\pm0.14}$  & & $\mathbf{1.55\pm0.03}$ & $\mathbf{2.86\pm0.05}$  &&  $\mathbf{2.22\pm0.08}$ & $\mathbf{4.32\pm0.09}$ && $\mathbf{2.98\pm0.08}$ & $\mathbf{5.10\pm0.15}$   \\
\midrule

SASRec    & $2.38\pm0.14$  & $4.49\pm0.21$  & & $1.60\pm0.07$ & $3.24\pm0.12$  &&  $2.26\pm0.11$ & $4.41\pm0.19$ && $3.02\pm0.15$ & $5.77\pm0.19$   \\

\it{w/} Random & $2.25\pm0.15$  & $4.03\pm0.26$  & & $1.61\pm0.06$ & $3.13\pm0.14$  &&  $2.34\pm0.13$ & $4.48\pm0.18$ && $3.05\pm0.15$ & $5.68\pm0.21$   \\

\it{w/} Focused & $2.47\pm0.11$  & $4.73\pm0.18$  & & $1.66\pm0.04$ & $3.32\pm0.09$  &&  $2.45\pm0.10$ & $4.83\pm0.14$ && $3.02\pm0.12$ & $5.47\pm0.17$   \\

\it{w/} Adversarial & $2.35\pm0.16$  & $4.37\pm0.21$  & & $1.62\pm0.03$ & $3.12\pm0.10$  &&  $2.43\pm0.11$ & $4.54\pm0.16$ && $3.15\pm0.10$ & $5.93\pm0.21$   \\

\it{w/} L2Aug & $\mathbf{2.61\pm0.12}$  & $\mathbf{4.87\pm0.19}$  & & $\mathbf{1.69\pm0.03}$ & $\mathbf{3.38\pm0.08}$  &&  $\mathbf{2.44\pm0.08}$ & $\mathbf{4.53\pm0.11}$ && $\mathbf{3.32\pm0.11}$ & $\mathbf{5.86\pm0.15}$   \\

\bottomrule
\end{tabular}}

\label{table:casual_results}
\end{table*}

\smallskip
\noindent\textbf{Baselines.} As the proposed method is model agnostic,  we apply it to various sequential recommendation models and compare its performance with other model-agnostic treatment methods to examine its effectiveness. 
We select three major \textit{sequential recommendation models} -- \textbf{GRU4Rec} \cite{hidasi2015session}, \textbf{SASRec} \cite{kang2018self} and \textbf{NextItNet} \cite{yuan2019simple}, which are built on top of Gated Recurrent Units (GRU), self-attention layers and stacked 1D dilated convolutional layers to capture the sequential patterns in user interaction sequences respectively. They are widely used in many applications and serve as the foundation for many advanced recommender systems.

Since there are no previous works focusing on improving casual user recommendations, for each sequential recommendation model, we compare the proposed \textit{L2Aug} with the following \textit{treatment methods} which are proved to alleviate the performance gap between different user groups. 
\squishlist
\item \textbf{Random}: It randomly drops the interactions of core users to obtain the synthetic data, which are combined with the original data (both core \& casual users) for training the recommender.
\item \textbf{Focused Learning} \cite{beutel2017beyond}: It treats the casual users $\mathcal{U}_{casual}$ as the focused set and performs a grid search for the best performing hyperparameters (i.e., the regularization) for improving recommendation accuracy on the focused set.
\item \textbf{Adversarial Reweighting} \cite{lahoti2020fairness}: It plays a minimax game between a recommender and an adversary. The adversary would adversarially assign higher weights to regions where the recommender makes significant errors, in order to improve the recommender's performance in these regions.
\squishend

\begin{figure*}[!t]
\vspace{-0.1in}
    \graphicspath{{figures/}}
    \centering
    \scalebox{0.99}{
    \subfigure[\textbf{Amazon\_CDs}]
    {
    \hspace{-0.1cm}
    \includegraphics[width=0.24\textwidth]{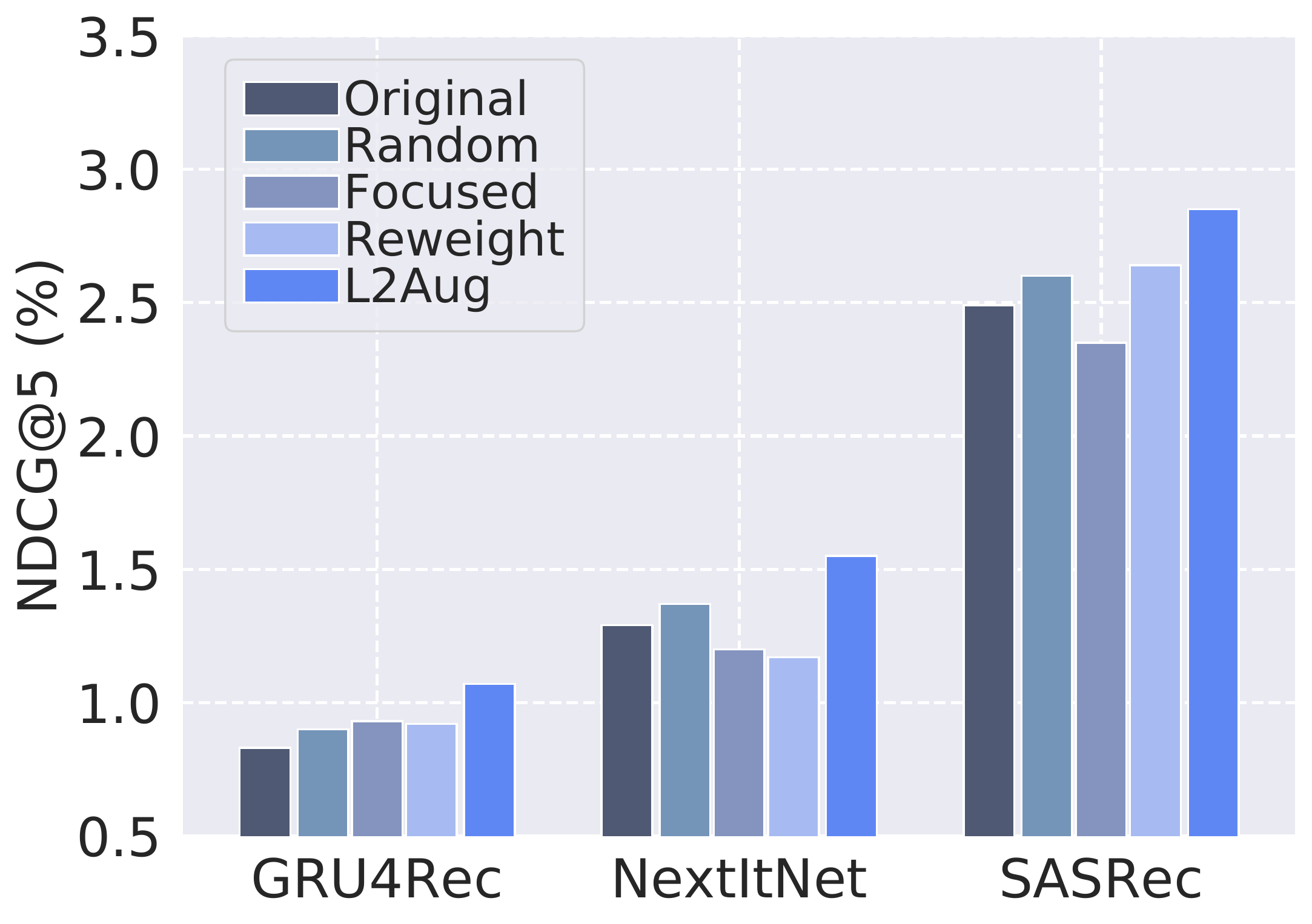}
    }
    \hspace{-0.1cm}
    \subfigure[\textbf{Amazon\_Books}]
    {
    \includegraphics[width=0.24\textwidth]{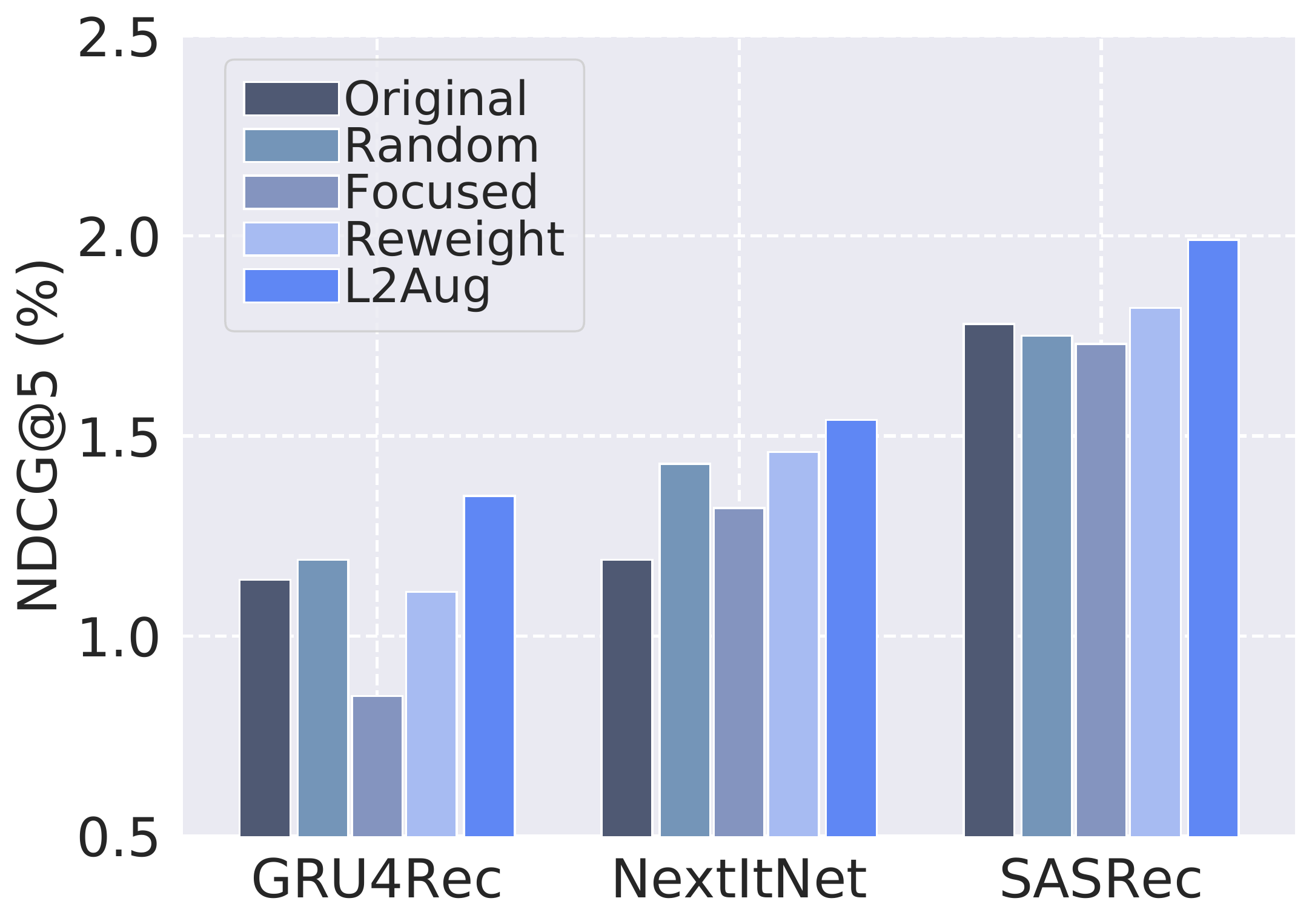}
    }
    \hspace{-0.1cm}
    \subfigure[\textbf{Amazon\_Movies}] 
    {
    \includegraphics[width=0.24\textwidth]{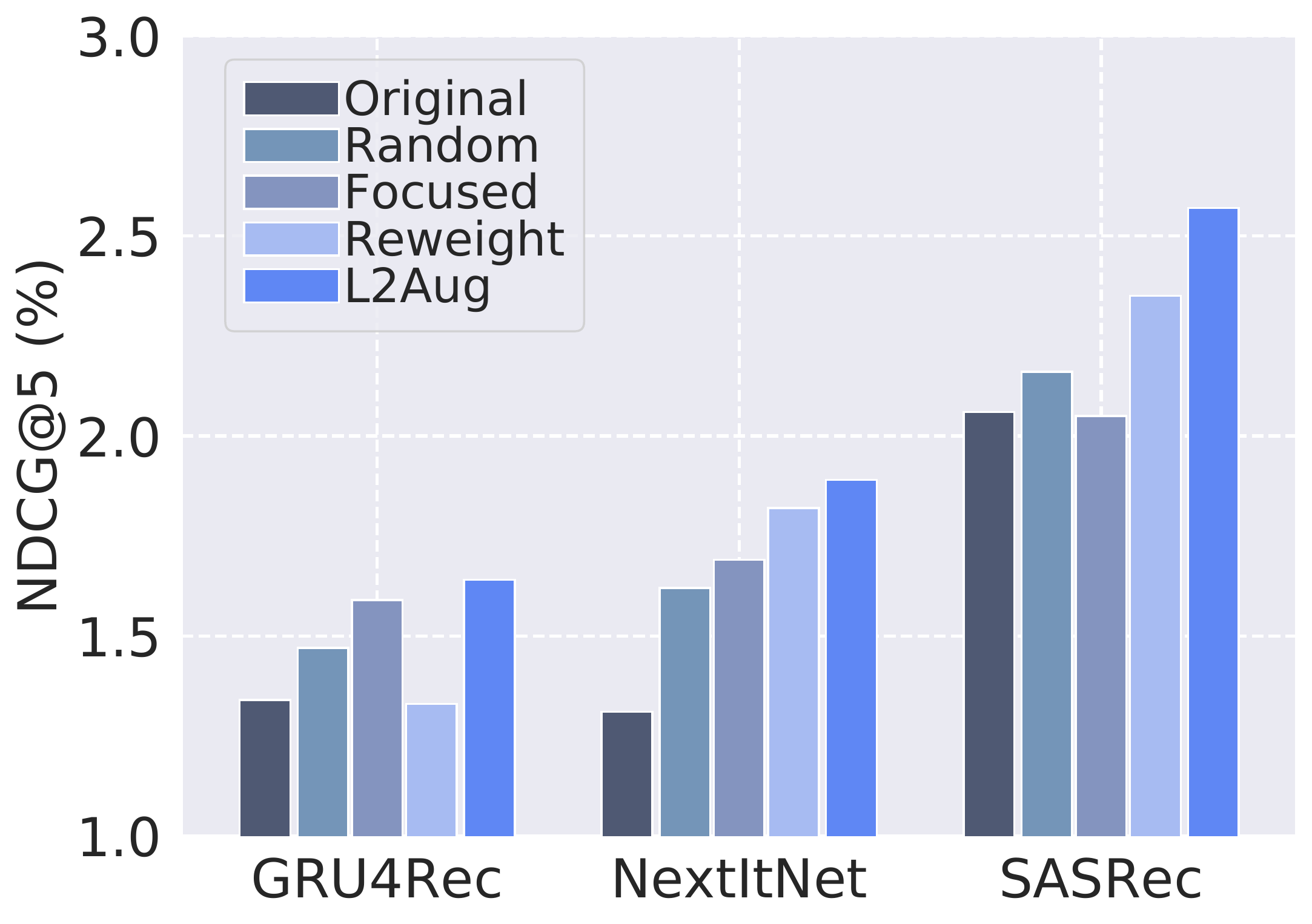}
    }
    \hspace{-0.1cm}
    \subfigure[\textbf{Goodreads}]
    {
    \includegraphics[width=0.24\textwidth]{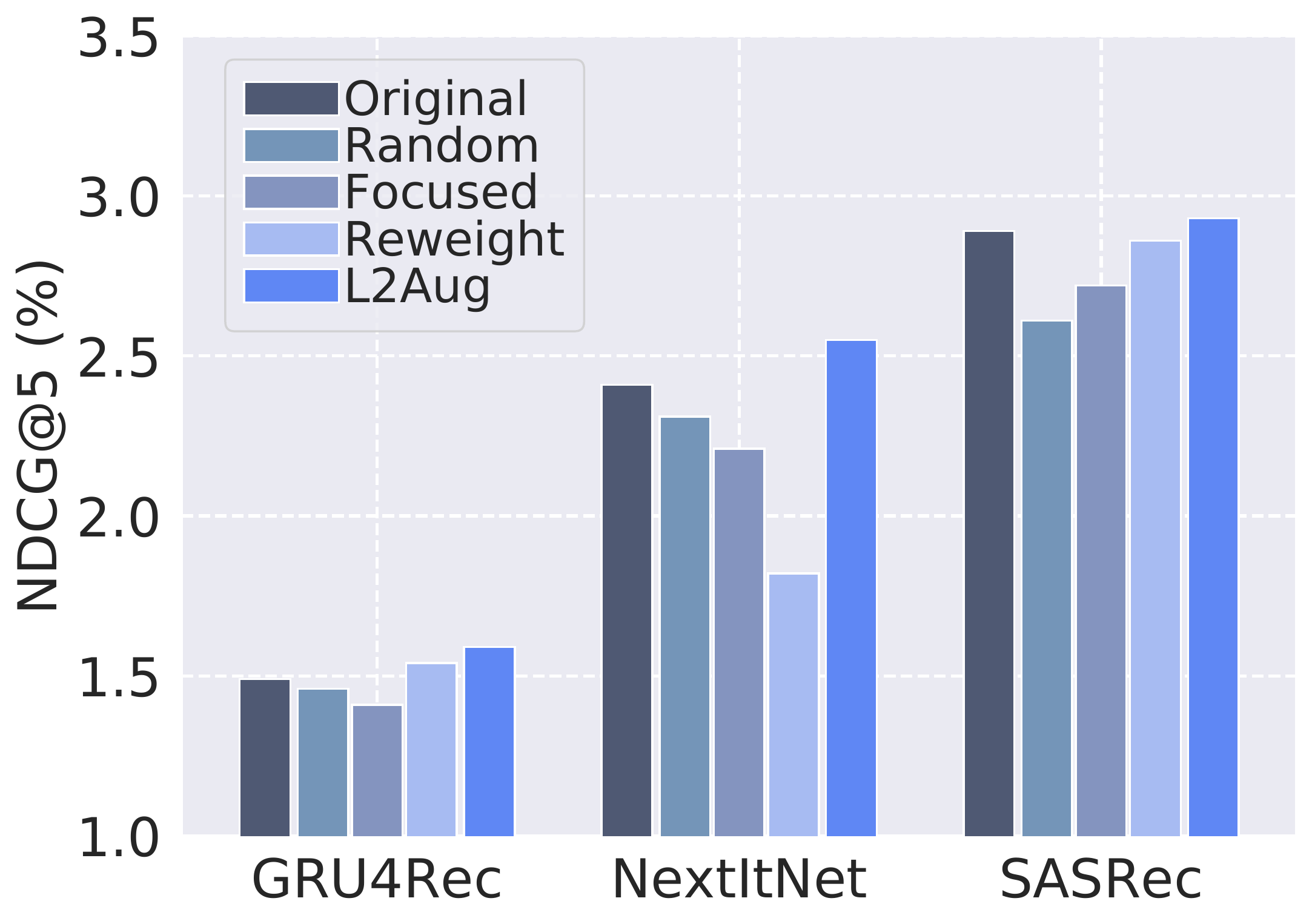}
    }}
     \vspace{-0.1in}
    \caption{Performance on \underline{core} user recommendation of various models on different datasets.}%
    \label{fig:N}
\end{figure*}

\smallskip
\noindent\textbf{Evaluation Metrics.} We train a recommendation model on the training data and then test it on the test data, where each of the users has one ground-truth item. We adopt the widely used Top-K metrics, including Hit Rate (HT@$K$) and Normalized Discounted Cumulative Gain (NDCG@$K$) to evaluate the recommendation performance. For user $u$, based on the scores $\hat{\textbf{y}}_u$ from Eq. (\ref{equ:inference}), we obtain the ranking $r_u$ of the ground-truth item. Hit rate indicates whether the ground-truth item appears in the Top-K list, i.e., $\mathrm{HT}_u@K = 1$ if $r_u \le K$ and $\mathrm{HT}_u@K = 0$ otherwise. Note that hit rate is equal to recall in this case since there is only one item in the test set for each user. When computing NDCG, the Ideal Discounted Cumulative Gain (IDCG) is equal to 1 for all users. Therefore, we have $\mathrm{NDCG}_u@K = \frac{1}{\log_2{(1+r_u)}}$ if $r_u \le K$ and $\mathrm{NDCG}_u@K = 0$ otherwise. In the following, we report the results averaged over all users in the test set with $K=5$ and $10$.

\smallskip
\noindent\textbf{Implementation Details.} 
All of the recommendation models are implemented in TensorFlow and optimized with Adam \cite{kingma2014adam}. We utilized the code published by or suggested by authors of the original work and keep the settings unchanged. We set the maximum sequence length to 200 and the batch size to 512 for all the datasets. For a fair comparison, the item embedding dimension is 100 and the negative sampling rate is set to 1 in the training process for all the models. We also implement our data augmentor in TensorFlow and adopt Adam as the optimizer. In Table \ref{table:casual_results}, we report the results based on the setting chosen on the validation set. In \textit{L2Aug}, 10\% of casual users among the training set are used to construct the meta validation set to compute reward for learning the data augmentation policy. The update frequency $\digamma$ is 5 for all the datasets.
More details for parameter selection and model implementation can be found in Appendices~\ref{sec:app_implementation} and \ref{sec:para_sen}.


\subsection{Overall Performance}


We conduct the experiments on the four datasets and run each experiment 10 times. The mean and variance of each metric are reported. We present quantitative results on recommendation performance metrics for both casual and core users, along with qualitative analyses, to showcase the efficacy of the proposed method.  

\subsubsection{Casual User Recommendation} 
We summarize the average performance of different baseline methods on all of the datasets in Table \ref{table:casual_results}. When combined with various sequential models, the proposed L2Aug outperforms all of the other treatment methods and achieves the best recommendation for casual users. In the following, we present more in-depth observations and analyses: 

\squishlist
\item  The simplest treatment of \textit{randomly} dropping part of the interactions from core users helps to improve casual user recommendations compared to the model trained on original data. This observation verifies the hypothesis that data augmentation can help bridge the gap between core and casual users.
\item By learning a recommendation model with a special focus on casual users, the \textit{focused learning} treatment can help improve model performance on casual users. Meanwhile, since recommenders tend to make inaccurate predictions on casual users,
\textit{adversarial reweighting} can guide the recommender to improve its performance on casual users, leading to more accurate recommendations for them. 
\item In general, the proposed L2Aug significantly outperforms all the baseline treatments on improving casual user recommendations for various widely-used sequential recommendation models. Take the \textit{Amazon\_CDs} dataset as an example, we find that L2Aug achieves 9.59\%, 6.58\%, and 9.90\% improvements for NDCG@5 with GRU, NextItNet and SASRec, respectively, compared against the best performing baseline treatment. We can conclude that L2Aug is effective in solving the challenging problem of improving casual user recommendations.
\squishend

\subsubsection{Core User Recommendation}
Besides the performance on casual users, we also report the recommendation performance on core users in Figure \ref{fig:N}, measured by NDCG@5;
results by other metrics can be found in Appendix \ref{sec:addition}. 
Although
\textit{focused learning} improves the recommendation on casual users, it loses its prediction power on core users. The \textit{adversarial reweighting} treatment, 
aiming at improve challenging data samples rather than specific user groups,  improves core user recommendations in some cases, but not always.  In contrast, the proposed L2Aug improves core user recommendations with various sequential recommendation models and outperforms all the other baseline treatments. These results showcase its effectiveness in bridging the gap between recommendation for casual users and core users,  leading to overall recommendation improvement. 

\subsubsection{Qualitative Analyses of the Augmented Data} Figure \ref{fig:preliminary}(a) plots the synthetic sequential data (i.e., augmented users) produced by \textit{L2Aug} along with the casual and core users from the Amazon review dataset. One can see the distribution of the interest continuity for the augmented users lies between that of the casual and core users from the original dataset, suggesting \textit{L2Aug} successfully distilled more consistent patterns from core users while adapting to the patterns of casual users. 

\begin{figure}
\includegraphics[width=0.30\textwidth]{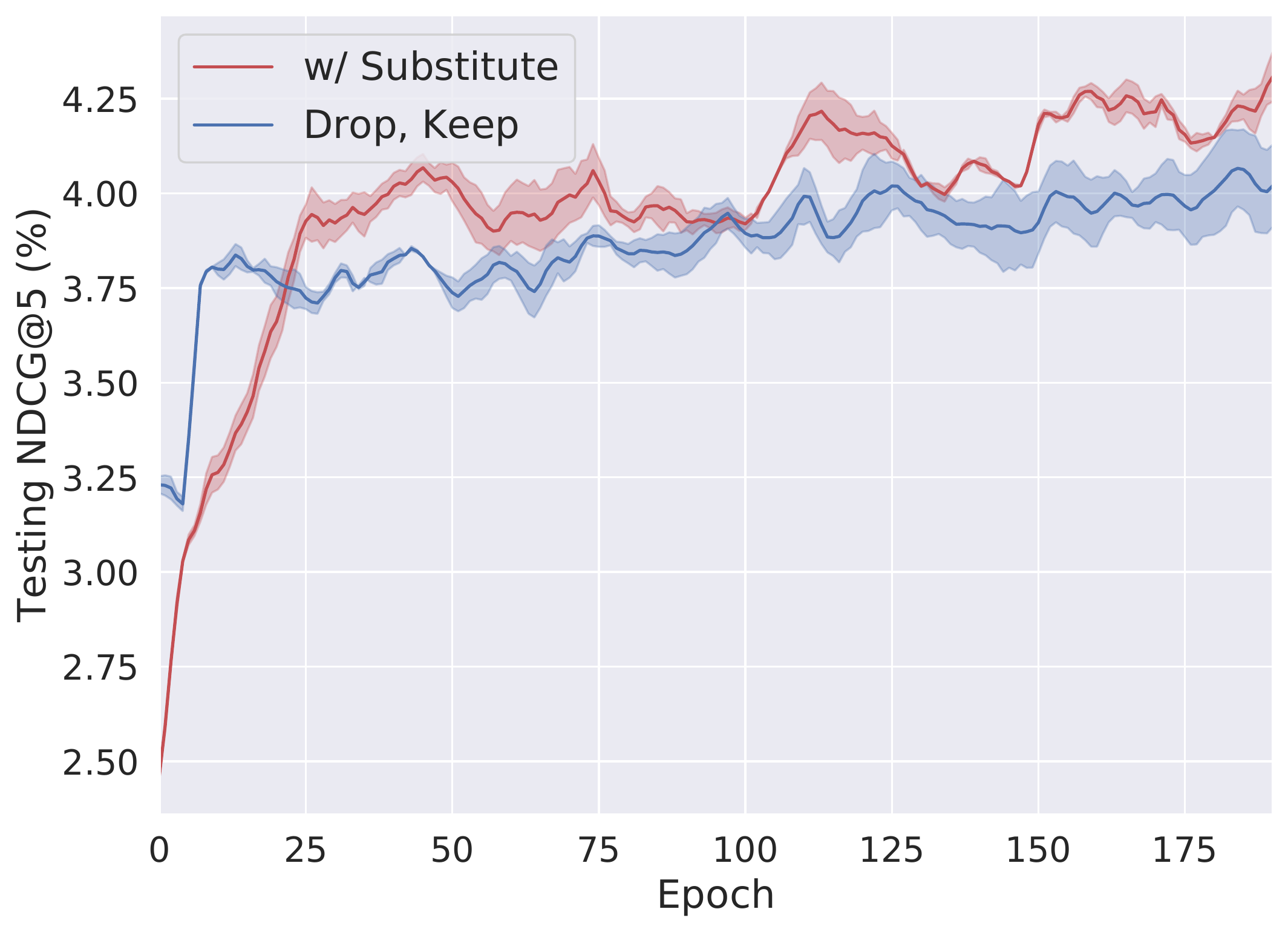}
\vspace{-0.05in}
\caption{L2Aug is easily extended to support more actions (i.e., substitute) with improved casual user recommendation.}
\vspace{-0.1in}
\label{fig:extend1}
\end{figure}

\subsection{Flexibility Analysis}

\subsubsection{Expanding the Action Space}
\label{sec:more_actions}

Recall that, so far, the augmentation policy can take two actions: ``keep'' and ``drop''. In this section, we examine the feasibility of expanding the action space of L2Aug, which makes the augmentor more versatile.
As an initial study, we consider the ``substitute'' action, which replaces an item with its most correlated item. 
We adopt the inverse user frequency (i.e., $|N(i)\cap N(j)|/\sqrt{|N(i)||N(j)|}$)  \cite{salton1983introduction,breese2013empirical} to define the correlation between two items (i.e., $i$ and $j$), in which $N(i)$ is the set of users interacted with item $i$. 
Figure \ref{fig:extend1} shows that adding the ``substitute'' action leads to higher recommendation performance on casual users; it also takes more epochs for the model to converge. Similarly, the proposed L2Aug can also be extended to support other actions like ``reorder'' and ``insert''.  Based on the observation, it can be conjectured that the proposed framework is capable of handling various editing actions for sequential data augmentation.

\subsubsection{Online Experiments}

\begin{figure}
\vspace{-0.05in}
    \centering
    \subfigure[List Size = 1] 
    {
    \includegraphics[width=0.47\columnwidth]{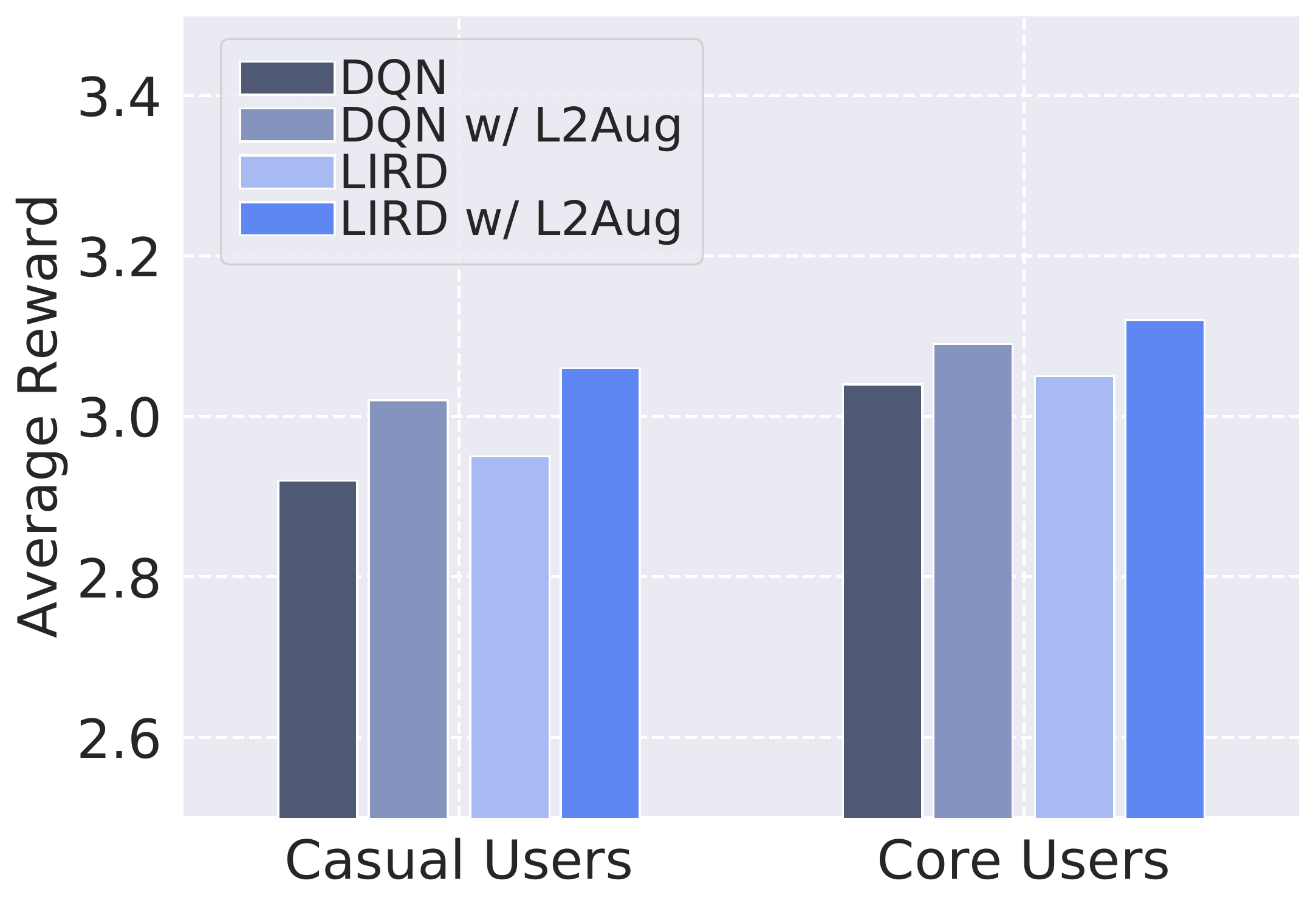}
    }
    \subfigure[List Size = 4]
    {
    \includegraphics[width=0.47\columnwidth]{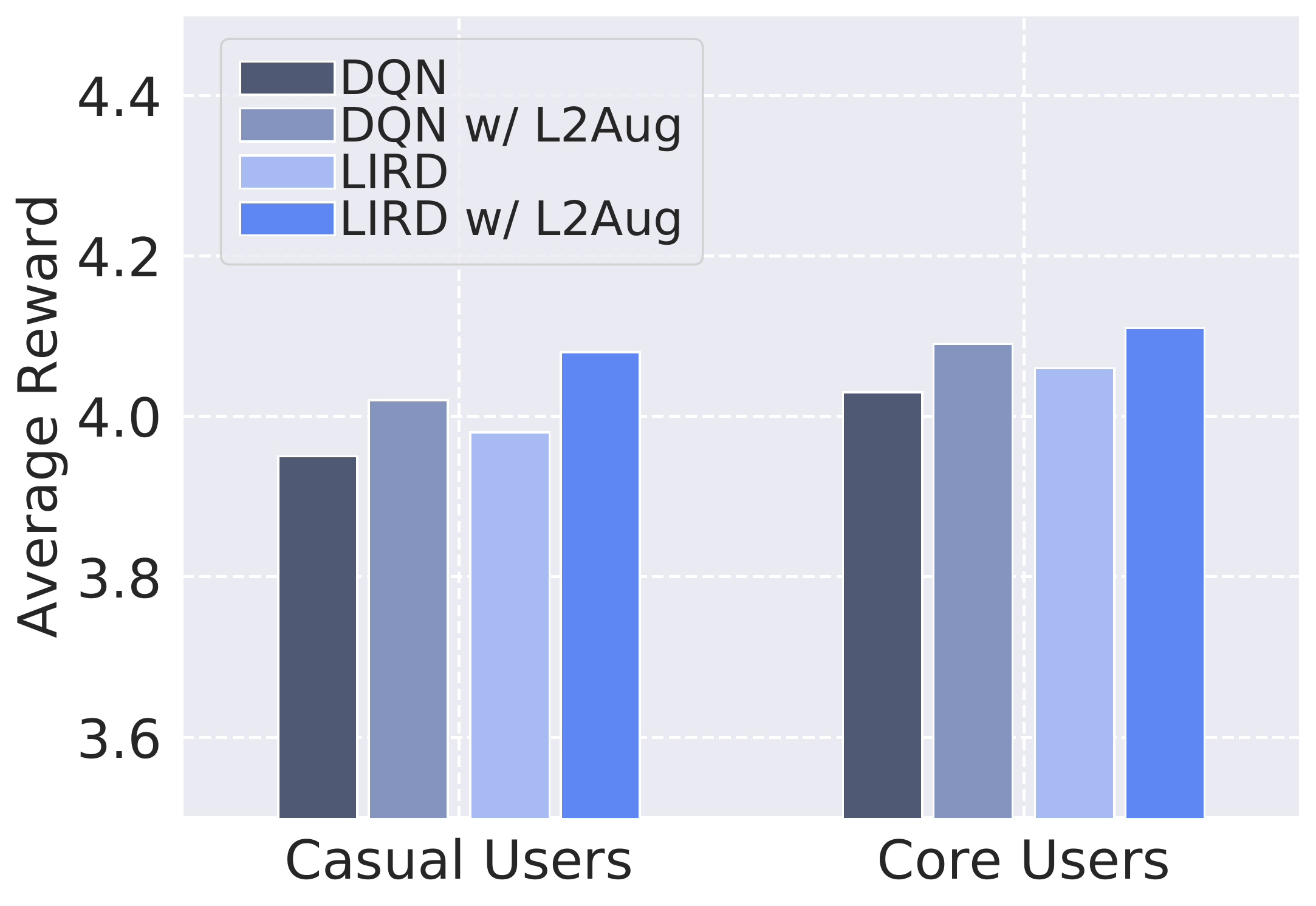}
    }
    \vspace{-0.05in}
    \caption{\textit{Online} test for List-wise Recommendation.}
    \label{fig:online}
    \vspace{-0.1in}
\end{figure}

The experiments so far are in offline settings, taking the observed user response on recommendations provided by the system as the ground-truth. 
Offline experiments have the drawback that we are not able to observe user response on counterfactual recommendations, i.e., items that were not shown to the users. To further evaluate the capability of the proposed model for real-world applications, we also conduct online experiments with simulation.  We follow \cite{zhao2017deep} to set up the online simulation environment. Given the user’s historical interactions and any recommendation candidate, it simulates the user response based on memory matching. 
This allows us to evaluate models on real-time responses (i.e., ratings) obtained from the simulator instead of relying on offline metrics. In the online experiment, we adopt the public MovieLen 100K dataset and split it into $7:3$ for training and testing, respectively. We treat users of the top 30\% visiting frequency as core users and the rest as casual users. In Figure \ref{fig:online}, DQN \cite{mnih2013playing} is the deep Q-learning method and LIRD \cite{zhao2017deep} is the state-of-the-art for list-wise recommendation. We use them as the recommenders (i.e., target model) in the L2Aug framework. Combined with L2Aug, both achieve improved performance on casual and core user recommendation under different list sizes, which further corroborates the efficacy of L2Aug under different recommendation scenarios.

\section{Related work}
\smallskip
\noindent\textbf{Sequential Recommendation.}
Instead of treating users as static, sequential recommendation aims to capture the sequential patterns from historical user interactions and infer the interesting items based on users' dynamic preferences. 
Early works propose to leverage Markov chains (MC) to model the transition among items and predict the subsequent interaction \cite{rendle2010factorizing,he2016fusing}. 
Grounded on the recent advances of deep learning techniques, 
there are lots of efforts on Recurrent Neural Networks (RNNs) to investigate users' sequential interactions~\cite{hidasi2015session,wu2017recurrent,wang2019recurrent,hidasi2018recurrent}. 
Meanwhile, Convolutional Neural Networks (CNN)-based recommenders also show superior performance. Caser \cite{tang2018personalized} is built on top of the 2D convolutional sequence embedding model and NextitNet \cite{yuan2019simple} investigates 1D CNNs with the dilated convolution filters for better performance. With its success in handling the textual data, self-attention layer (transformer) \cite{vaswani2017attention} is adopted in SASRec \cite{kang2018self} and Bert4Rec \cite{sun2019bert4rec} to generate dynamic user embedding based on their interaction sequences. More recently, Graph Neural Networks have been exploited in \cite{wu2019session,wang2020next,ma2020memory} to encode the contextual information for more accurate user modeling in sequential recommendation. 

\smallskip
\noindent\textbf{Long-tailed and Cold-User Recommendation.}
Several research topics in recommender systems are related to improving casual user recommendations.
\citet{beutel2017beyond} define the focused learning problem and propose to find different optimal solutions for different user groups through hyperparameter optimization and a customized matrix factorization objective. 
There are other efforts trying to learn multiple recommendation models and then select the best performing one from the pool for each user \cite{luo2020metaselector,ekstrand2012recommenders}. 
Another line of research focuses on domain transfer and improving the recommendation for users who have very few observed interactions~\cite{wang2019minimax,wang2021sequential}. Most of these works rely on side information or contextual data \cite{lee2019melu,li2019zero,kang2019semi,bi2020dcdir}.
\citet{yin2020learning} propose to learn the transferable parameters between data-rich head users and data-ill tail users, leading to an improved recommendation for both groups. Instead, we aim to bridge the gap between users who visit more frequently and users who casually visit the service via data augmentation. 


\smallskip
\noindent\textbf{Data Augmentation.}
Data augmentation is a widely adopted strategy for training deep neural models in computer vision \cite{cubuk2019autoaugment,krizhevsky2012imagenet}, graph learning \cite{ding2022data} and natural language processing \cite{wei2019eda,kobayashi2018contextual}. The idea is to generate more diverse data samples to improve model robustness. While most of the early work manually designs  dataset-specific augmentation strategies or implementations, recent attention has been paid to automating the process of generating effective augmentation for the target dataset. Generative Adversarial Networks (GANs) have been commonly used to generate additional data by learning the distribution of the training data under the MinMax framework \cite{gao2021recommender,chae2019rating,antoniou2017data,croce2020gan}. 
As an alternative to GAN-based methods, AutoAugment \cite{cubuk2019autoaugment} automatically searches for the best augmentation policy using reinforcement learning. However, there is still a research gap between generating additional images or textual data and augmenting sequential user interaction data; the latter is still understudied.
Recently, CL4Rec \cite{xie2021contrastive} proposes to construct different views of an interaction sequence with simple data augmentation using the contrastive loss, resulting in a more robust recommendation system. We focus on learning to generate effective augmented interaction sequences that mimic the patterns of casual users while inheriting informative transition patterns of core users. 

\section{Conclusion and Future Work}
Users who come to the online platform are heterogeneous in activity levels. To bridge the gap between recommendation for core users and casual users, we propose a model-agnostic framework \textit{L2Aug} to learn the data augmentation policy and improve the recommendation system with the generated data. With experiments on four real-world public datasets, the proposed \textit{L2Aug} can outperform other treatment methods and improve casual user recommendation without sacrificing the recommendation for core users. Furthermore, \textit{L2Aug} is flexible in supporting multiple augmentation actions and different experimental (i.e., offline and online) setups.
For future work, we are interested in extending such a ``learning to augment'' concept to other application scenarios (e.g., cross-domain, cold-start) for improving the robustness and adaptivity of different recommendation systems. Moreover, we are interested in extending our augmentation policy from the bandit setup studied here to the reinforcement learning setup, where the agent chooses editing actions depending on its previous decisions.

\newpage
\balance
\bibliographystyle{ACM-Reference-Format}
\bibliography{sample-bibliography} 

\newpage
\appendix
\section{Appendix}
\subsection{Datasets Details}
\label{sec:preprocess}

\smallskip
\noindent\textbf{Amazon.} This is a public dataset\footnote{\url{http://jmcauley.ucsd.edu/data/amazon/links.html}} obtained from the E-commerce website -- Amazon, which includes users' reviews and ratings on products from May 1996 to July 2014. The dataset is split based on the product categories. In the experiments, we focus on three of the most popular categories to exploit the efficiency of the models in E-commerce scenarios.

\noindent\textbf{Gooreads.} The public dataset\footnote{\url{https://sites.google.com/eng.ucsd.edu/ucsdbookgraph/home}} were collected in late 2017 from Goodreads, which is a online platforms for users to share their opinions on books. In the experiments, we focus on users' reviews and ratings as their interactions with books. We treat different ratings and reviews equally and convert the interactions into binary feedback (i.e., interacted or not-interacted).

For each of the datasets, we group the interactions by users and then sort them based on their timestamps. Firstly, we calculate the average time gap between their consecutive interactions of each user. Then, we classify users with average time gap less than 30 days as core users, otherwise we label them as casual users. Furthermore, to avoid information leakage during the training process, we split each dataset with a corresponding cutting timestamp $T$, such that the interactions before $T$ are used for model training. And for each users, the first interaction after $T$ is for parameter fine-tuning and the second one is for testing. We select the cutting timestamp $T$ to be January 1, 2009 for all the datasets. 

\subsection{Implementation Details}
\label{sec:app_implementation}


\smallskip
\noindent\textbf{Implementation of Baselines.} 
For all the sequential recommendation models, we adopt their implementations in Tensorflow so that they are compatible with the treatment methods. 
\squishlist
\item \textbf{GRU4Rec} \cite{hidasi2015session}: We adopt the Tensorflow implementation\footnote{\url{https://github.com/Songweiping/GRU4Rec_TensorFlow}} and use one GRU layer with 100 hidden units as in the default setting.
\item \textbf{SASRec} \cite{kang2018self}: The implementation\footnote{\url{https://github.com/kang205/SASRec}} is published with the original paper. As suggested by the authors, we set the number of heads and the number of blocks in self-attention layer as 1 and 2, correspondingly.
\item \textbf{NextItNet} \cite{yuan2019simple}: We utilize the code\footnote{\url{https://github.com/fajieyuan/WSDM2019-nextitnet}} published by the authors and keep the settings.
unchanged. 
\squishend

As for the the baseline treatment methods, we combine them with various sequential recommendation models to examine its efficiency in improving casual user recommendation. 
\squishlist
\item \textbf{Random}: We randomly drops the interactions of core users to obtain the synthetic data sequences. The dropping rate is selected based on the performance on validation set.
\item \textbf{Focused Learning} \cite{beutel2017beyond}: Following the idea in the paper, we grid search for the best performing hyperparameters with the  hyperparameter optimization framework KerasTuner \cite{omalley2019kerastuner}\footnote{\url{https://github.com/keras-team/keras-tuner}}.
\item \textbf{Adversarial Reweighting} \cite{lahoti2020fairness}: We adapt the original implementations\footnote{\url{https://github.com/google-research/google-research/tree/master/group_agnostic_fairness}} to the setup of sequential recommendation via replacing the classifer with the recommendation models.

\squishend

Note that to achieve the best performance for each model, we perform a grid search for the dropout rate in \{0.2, 0.3, 0.4, 0.5, 0.6, 0.7, 0.8\} for \textit{Random}, the regularization weight $\lambda$ in \{$10^{-5}$, $10^{-4}$, $10^{-3}$, $10^{-2}$, $10^{-1}$\} for \textit{Focused Learning}, and the learning rate in \{$10^{-5}$, $10^{-4}$, $10^{-3}$, $10^{-2}$, $10^{-1}$\} for all methods.


\subsection{Parameter Sensitivity}
\label{sec:para_sen}
\begin{figure}[h]
\includegraphics[width=0.32\textwidth]{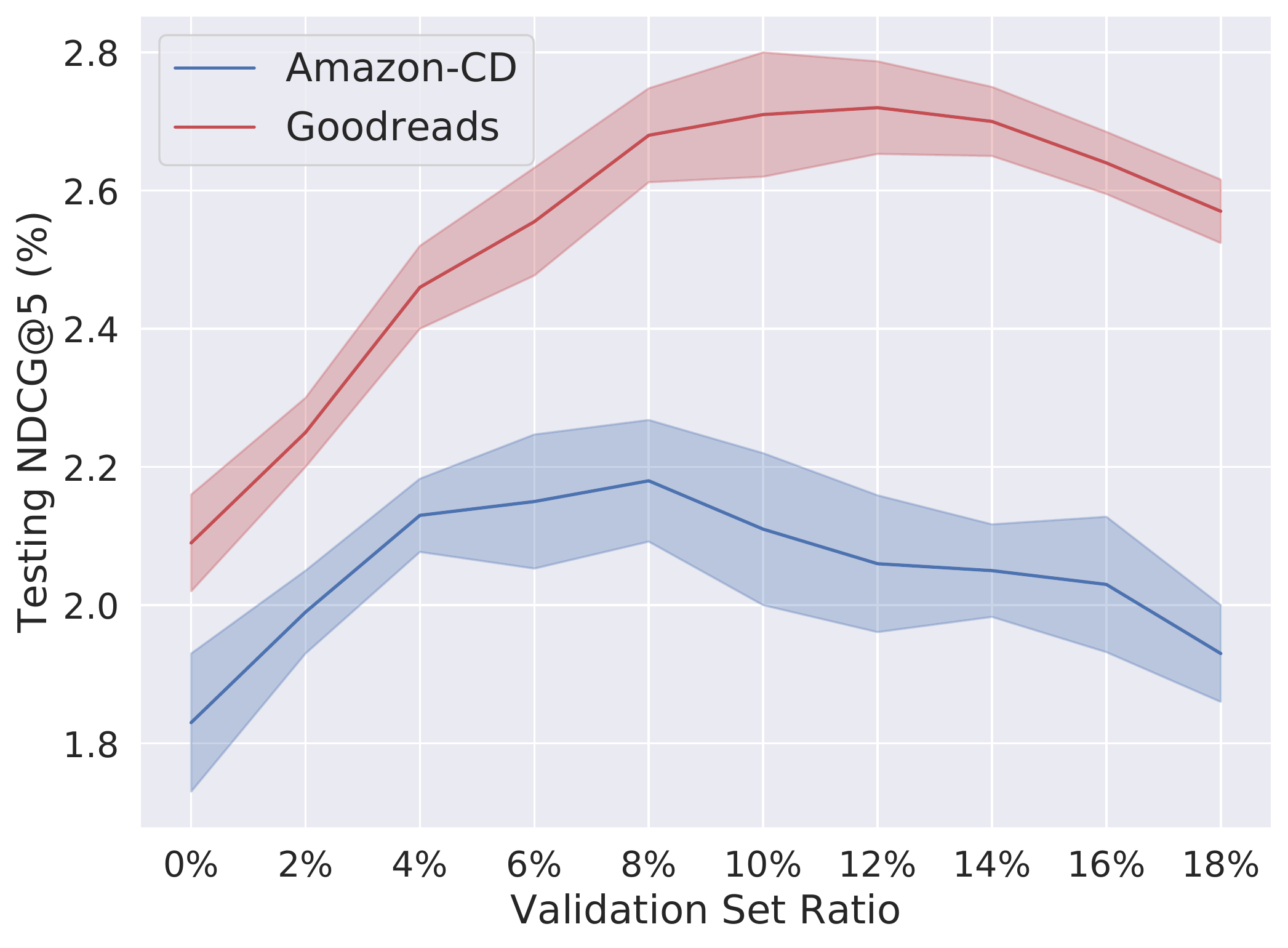}
\caption{Test performance VS Meta validation ratio}
\label{fig:extend2}
\end{figure}

One key design in L2Aug is to elicit the signal from meta validation set to guide the meta-learner for learning an effective augmentation policy. Note that the meta validation set is a small subset sampled from the casual user interaction sequences. In this section, we conduct experiments to evaluate the influence of the size of the meta validation set. In Figure \ref{fig:extend2}, we change the ratio of the meta validation set (i.e., $\frac{M}{N}$) and show the resulted recommendation performance on casual users with SASRec and L2Aug. We can observe that casual user recommendation would be improved with larger meta validation set, which demonstrates the effective guidance the meta validation set can provide to achieve our goal. However, sampling a meta validation set with size larger than $10\%$ would not bring in positive effect to the recommendation. Thus in the experiments, we select the ratio to be $10\%$ for the best performance.

\subsection{Additional Experimental Results}
\label{sec:addition}

\begin{figure*}
\vspace{-0.1in}
    \graphicspath{{figures/}}
    \centering
    \scalebox{0.99}{
    \subfigure[\textbf{Amazon\_CDs}]
    {
    \hspace{-0.1cm}
    \includegraphics[width=0.24\textwidth]{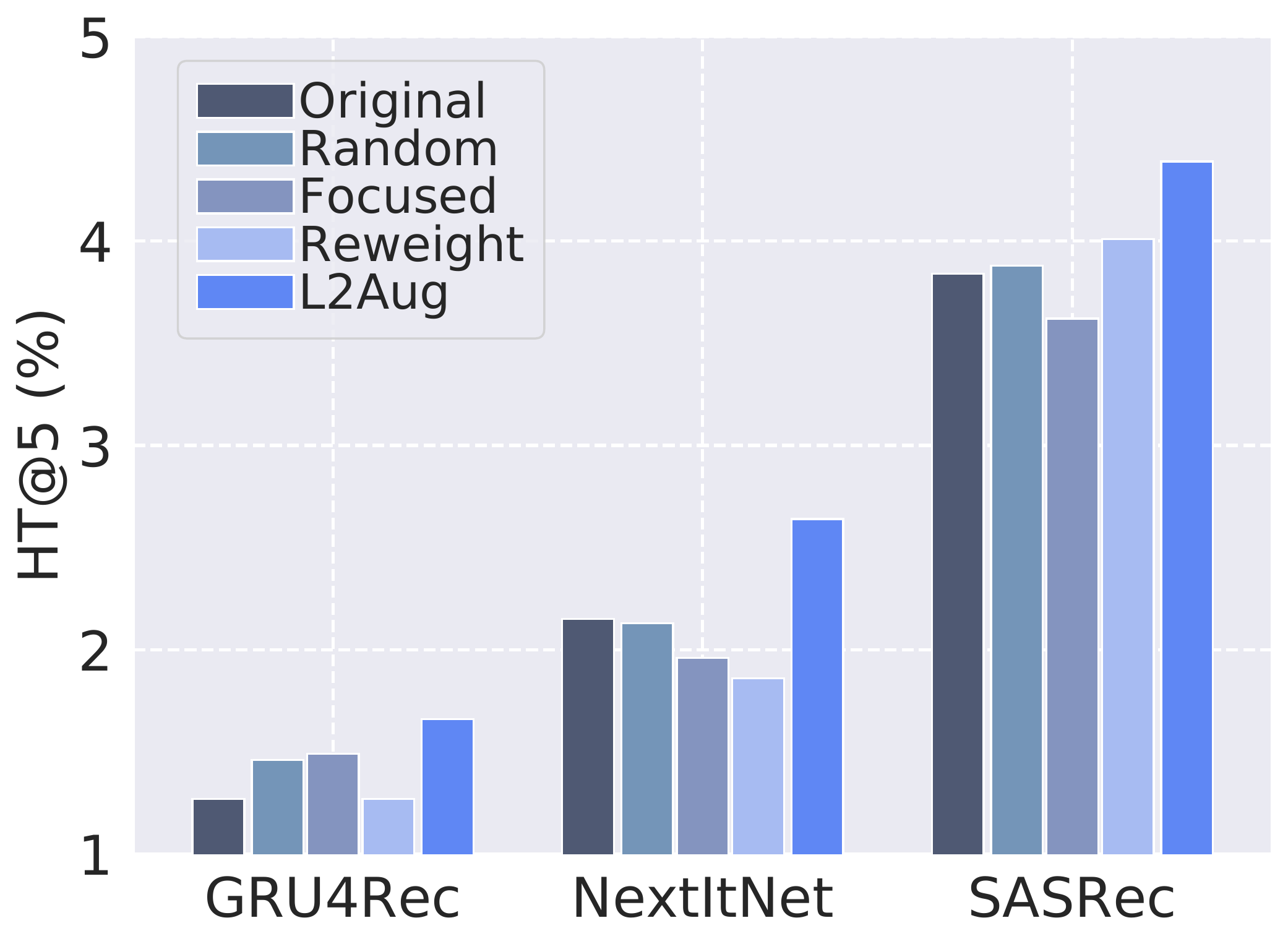}
    }
    \hspace{-0.1cm}
    \subfigure[\textbf{Amazon\_Books}]
    {
    \includegraphics[width=0.24\textwidth]{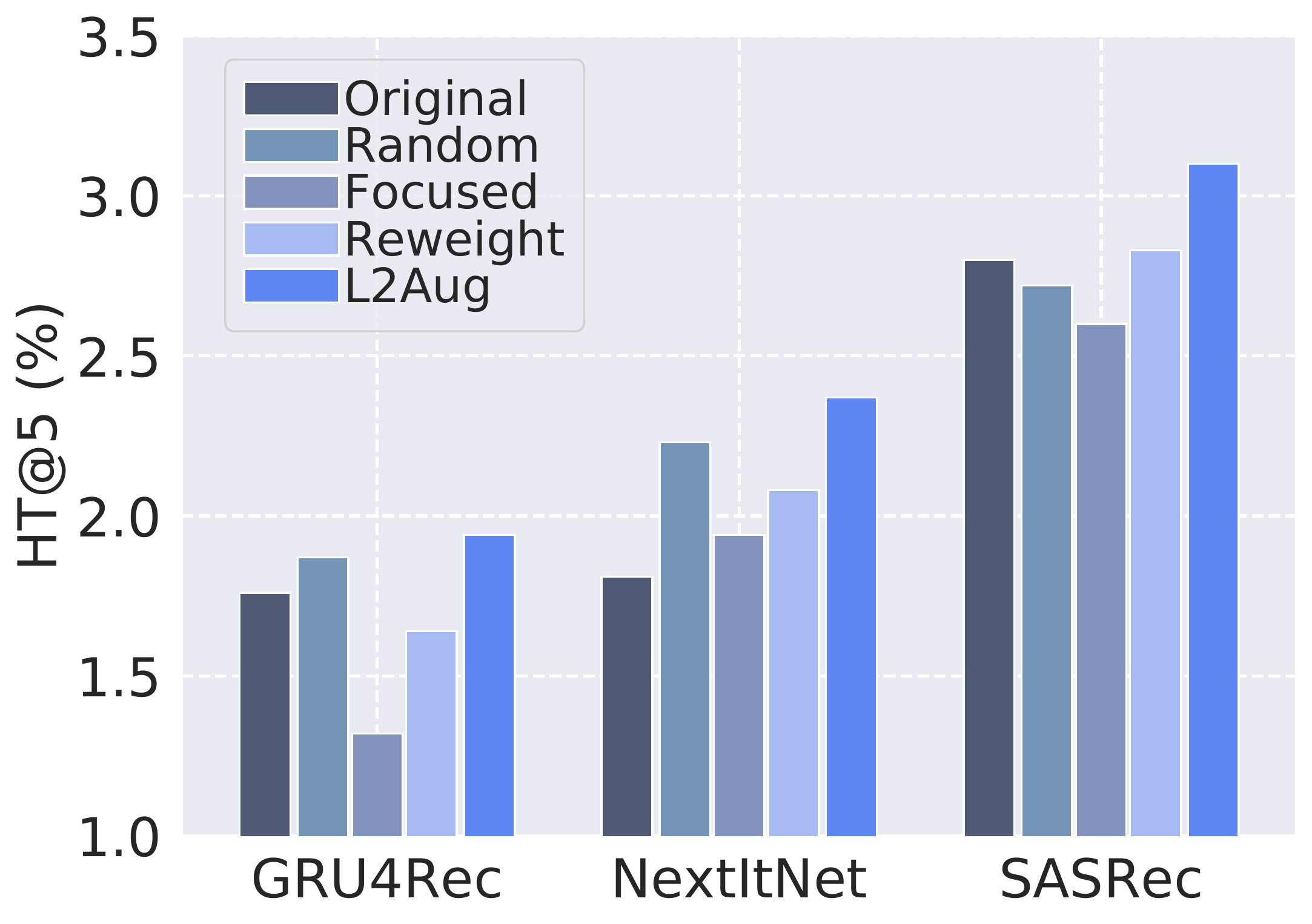}
    }
    \hspace{-0.1cm}
    \subfigure[\textbf{Amazon\_Movies}] 
    {
    \includegraphics[width=0.24\textwidth]{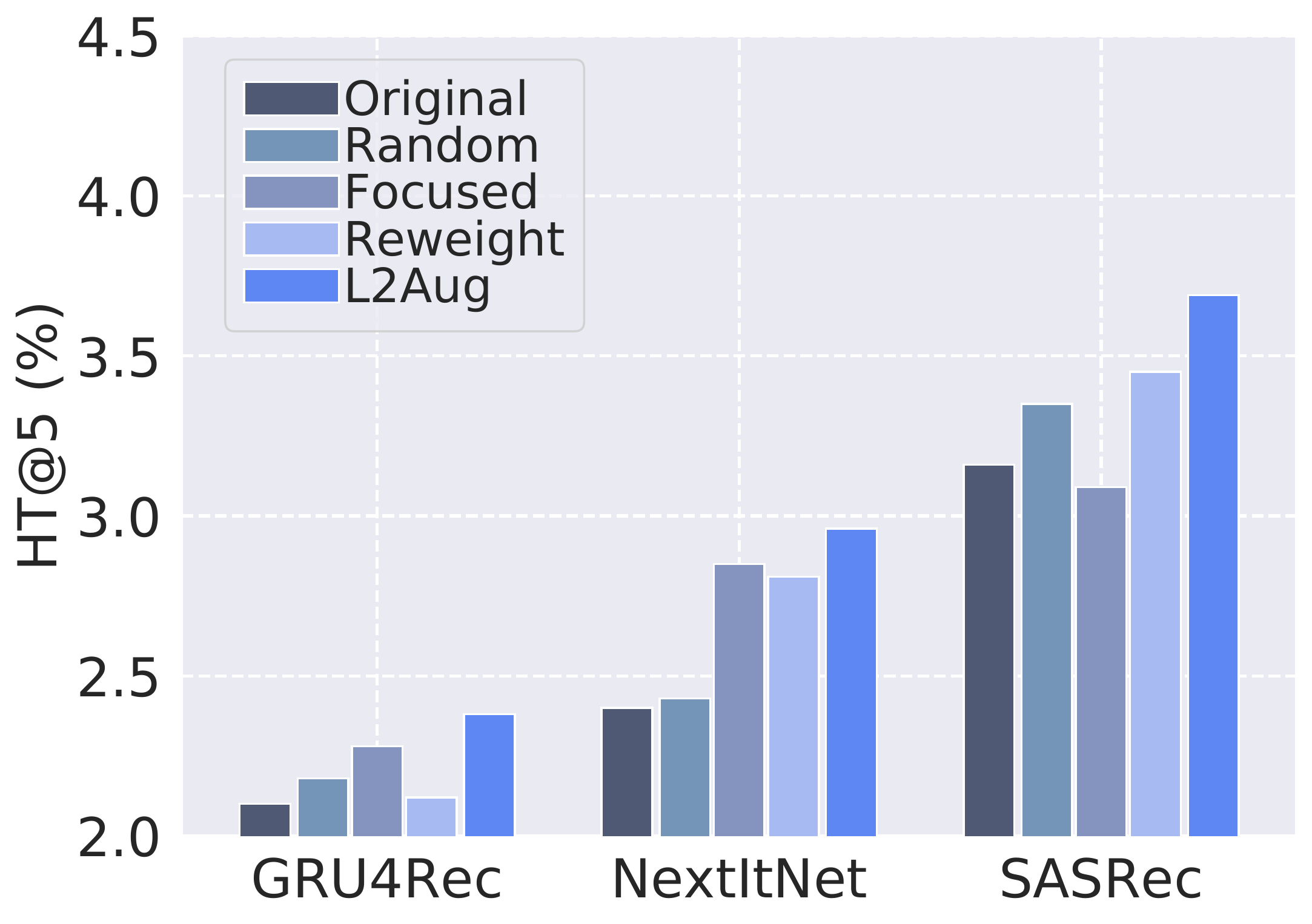}
    }
    \hspace{-0.1cm}
    \subfigure[\textbf{Goodreads}]
    {
    \includegraphics[width=0.24\textwidth]{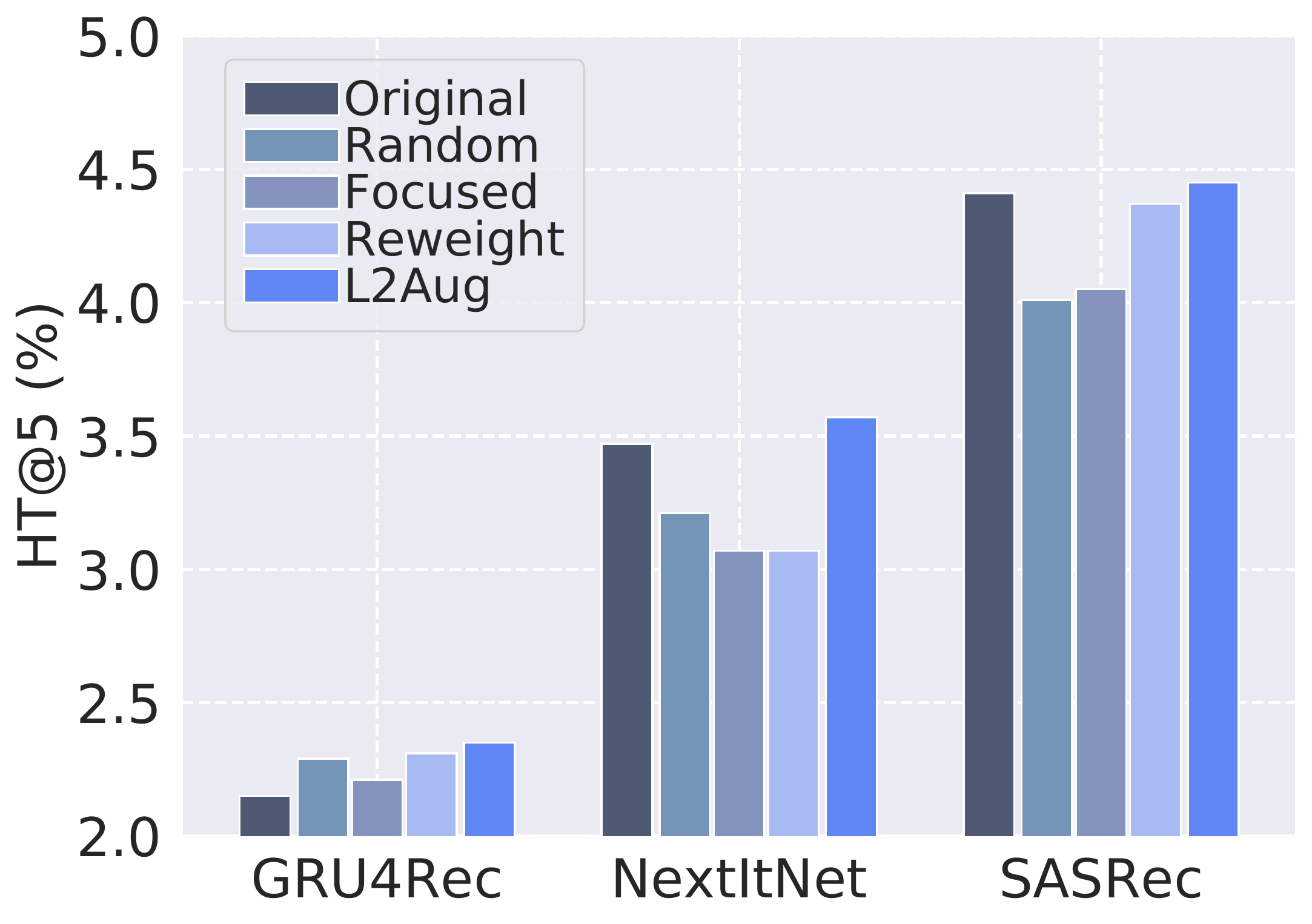}
    }}
     \vspace{-0.1cm}
    \caption{Performance (HT@5) in offline setup on Core User Recommendation of various models on different datasets.}%
    \label{fig:ht5}
\end{figure*}

\begin{figure*}[!b]
    \graphicspath{{figures/}}
    \centering
    \scalebox{0.99}{
    \subfigure[\textbf{Amazon\_CDs}]
    {
    \hspace{-0.1cm}
    \includegraphics[width=0.24\textwidth]{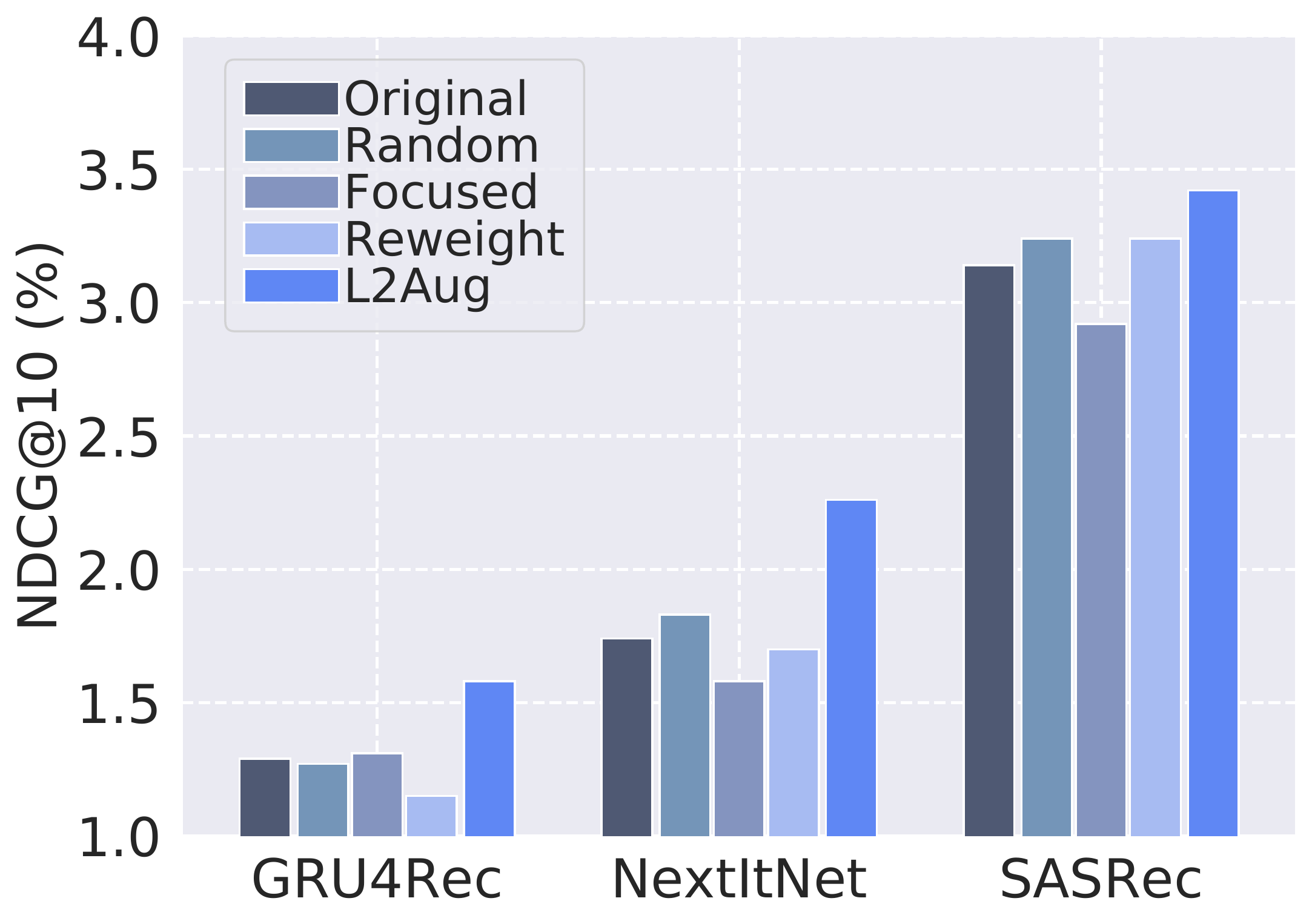}
    }
    \hspace{-0.1cm}
    \subfigure[\textbf{Amazon\_Books}]
    {
    \includegraphics[width=0.24\textwidth]{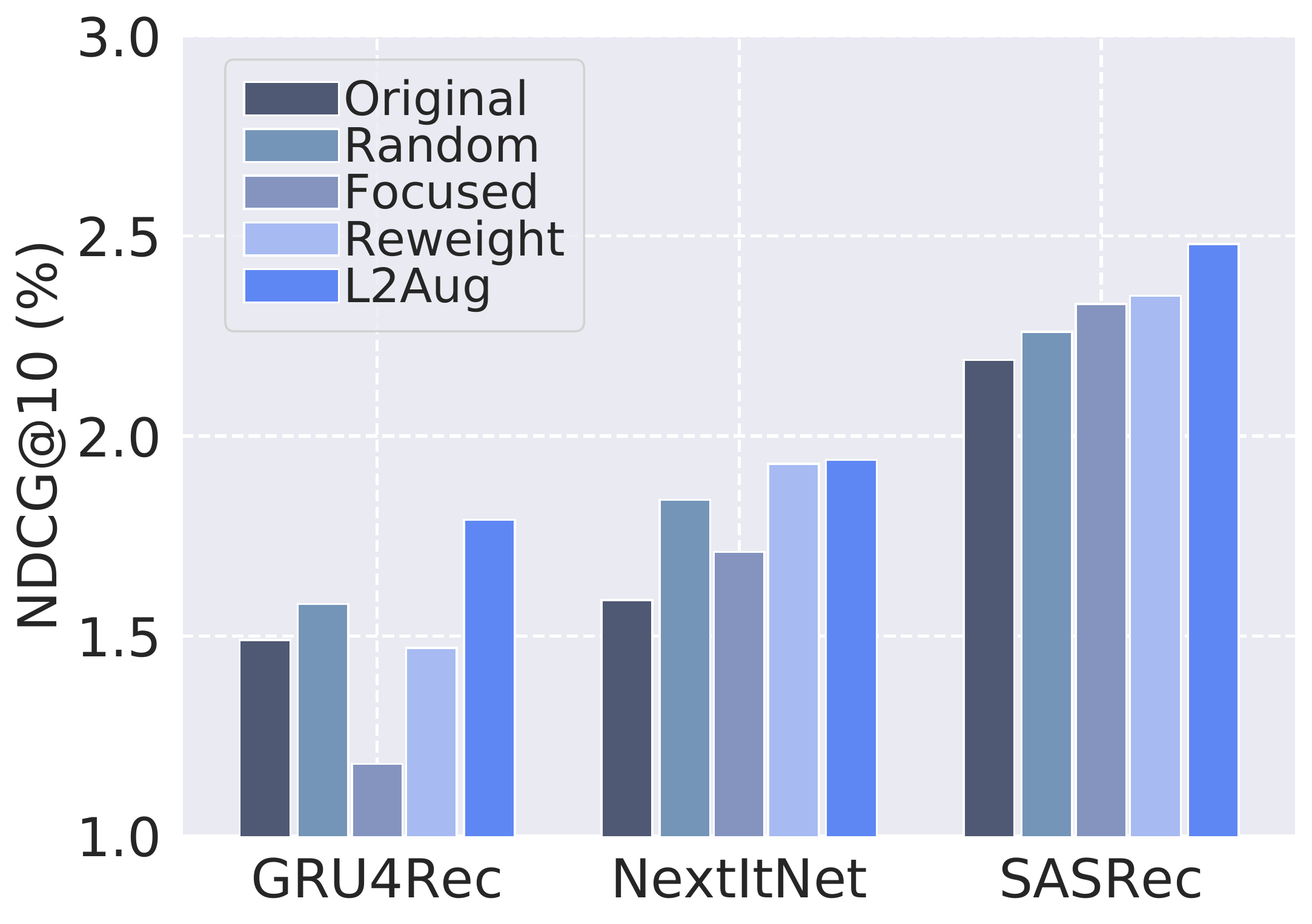}
    }
    \hspace{-0.1cm}
    \subfigure[\textbf{Amazon\_Movies}] 
    {
    \includegraphics[width=0.24\textwidth]{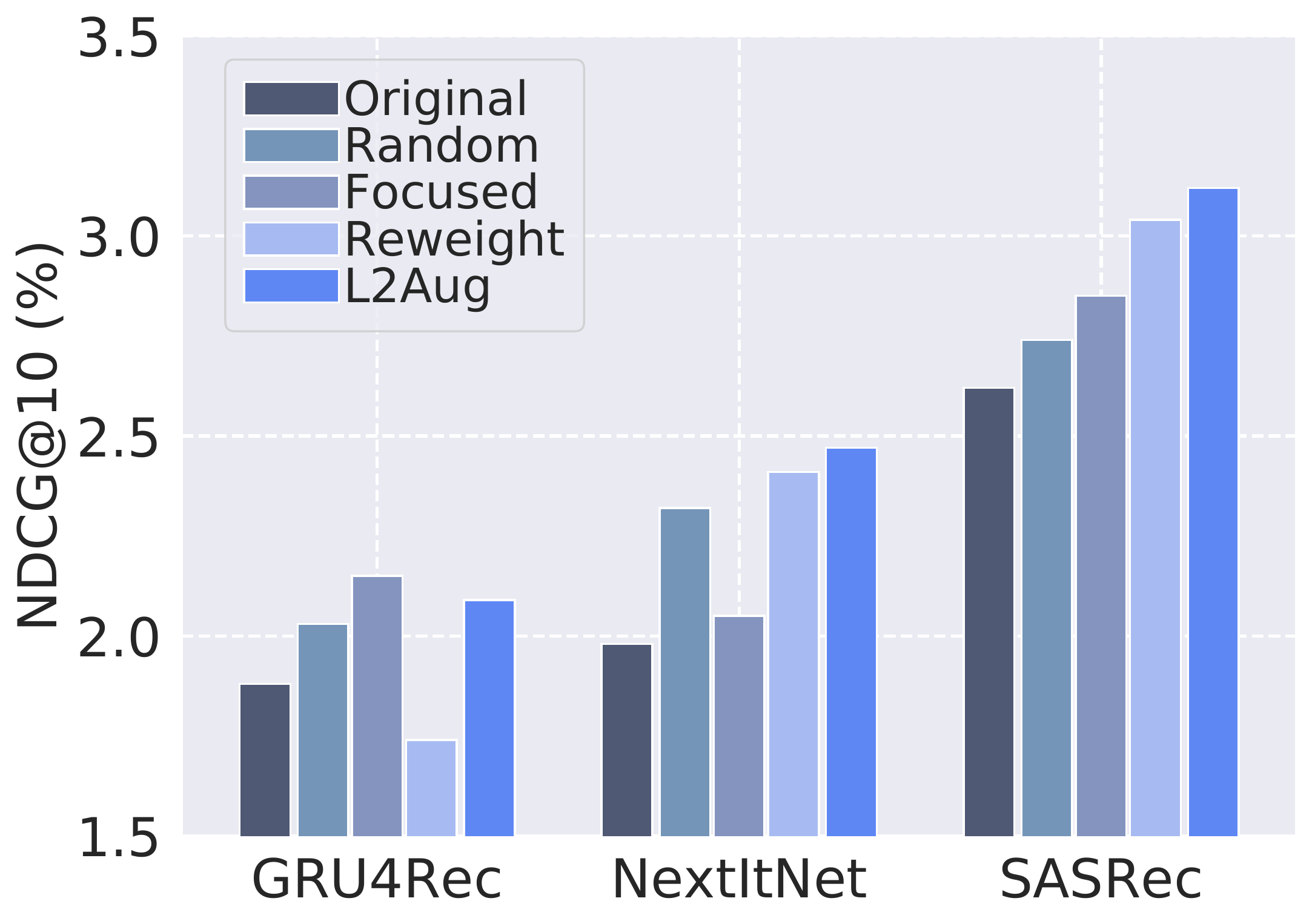}
    }
    \hspace{-0.1cm}
    \subfigure[\textbf{Goodreads}]
    {
    \includegraphics[width=0.24\textwidth]{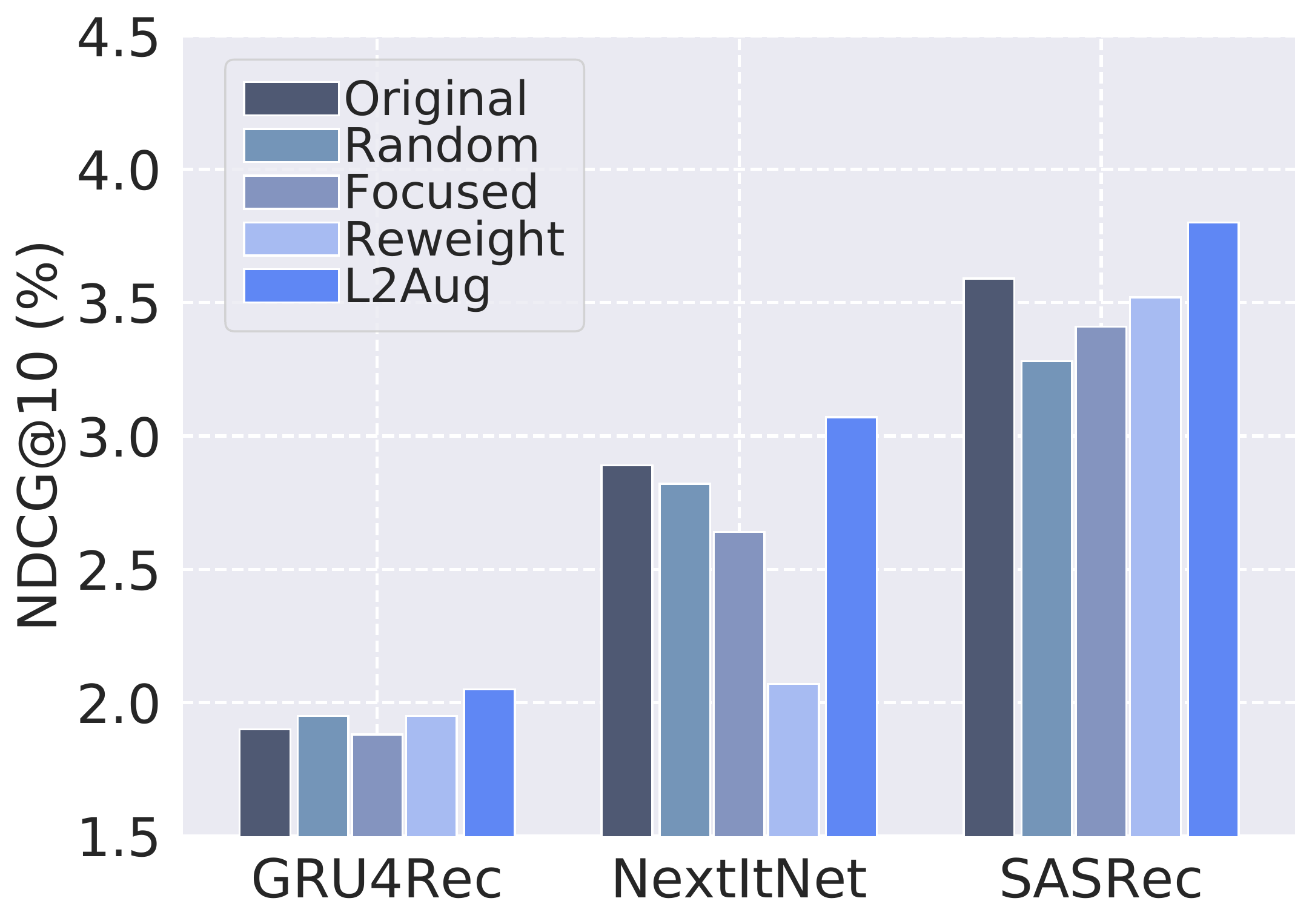}
    }}
     \vspace{-0.1cm}
    \caption{Performance (NDCG@10) in offline setup on Core User Recommendation of various models on different datasets.}%
    \label{fig:ndcg10}
\end{figure*}

\begin{figure*}[!b]
    \graphicspath{{figures/}}
    \centering
    \scalebox{0.99}{
    \subfigure[\textbf{Amazon\_CDs}]
    {
    \hspace{-0.1cm}
    \includegraphics[width=0.24\textwidth]{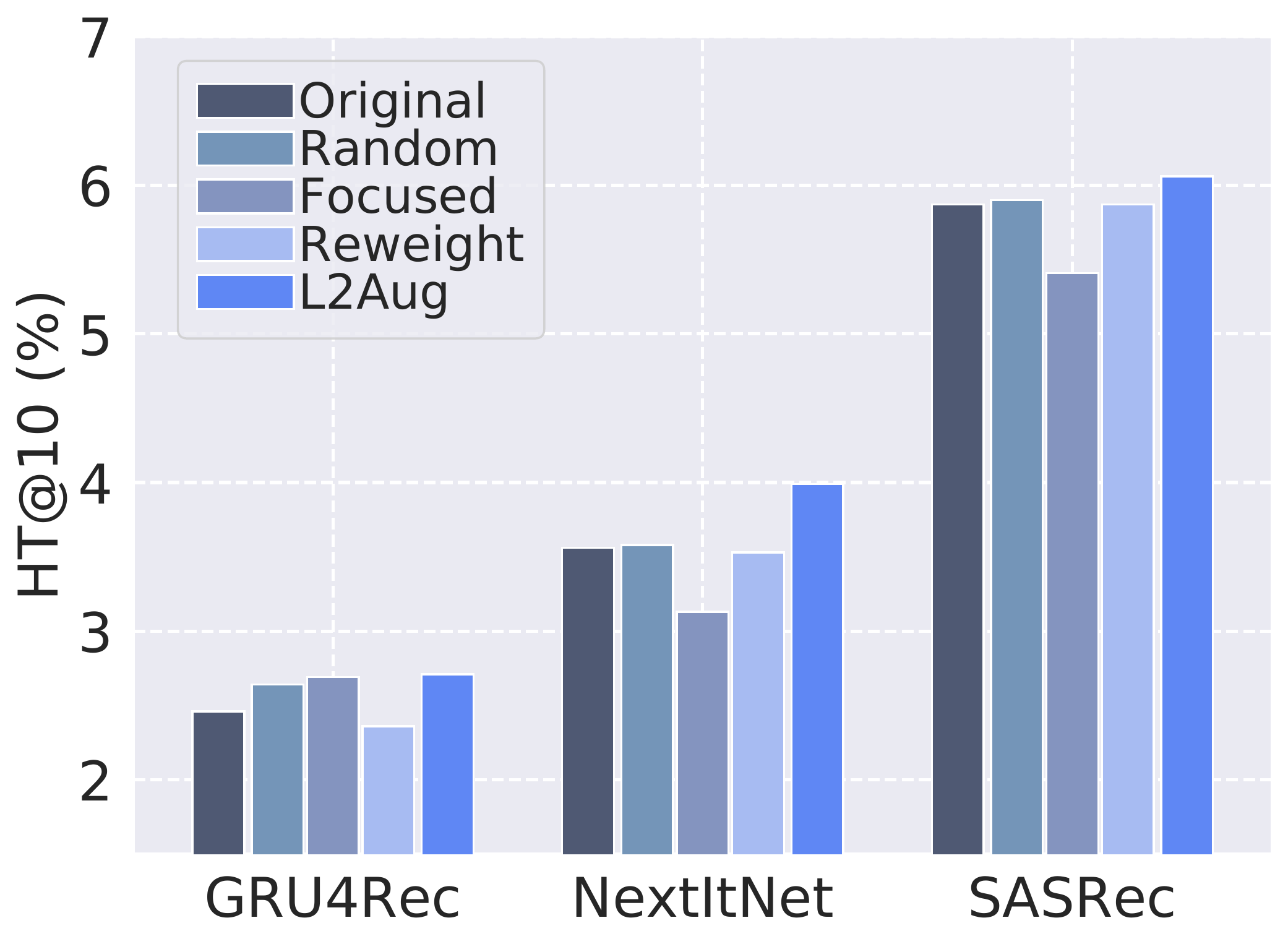}
    }
    \hspace{-0.1cm}
    \subfigure[\textbf{Amazon\_Books}]
    {
    \includegraphics[width=0.24\textwidth]{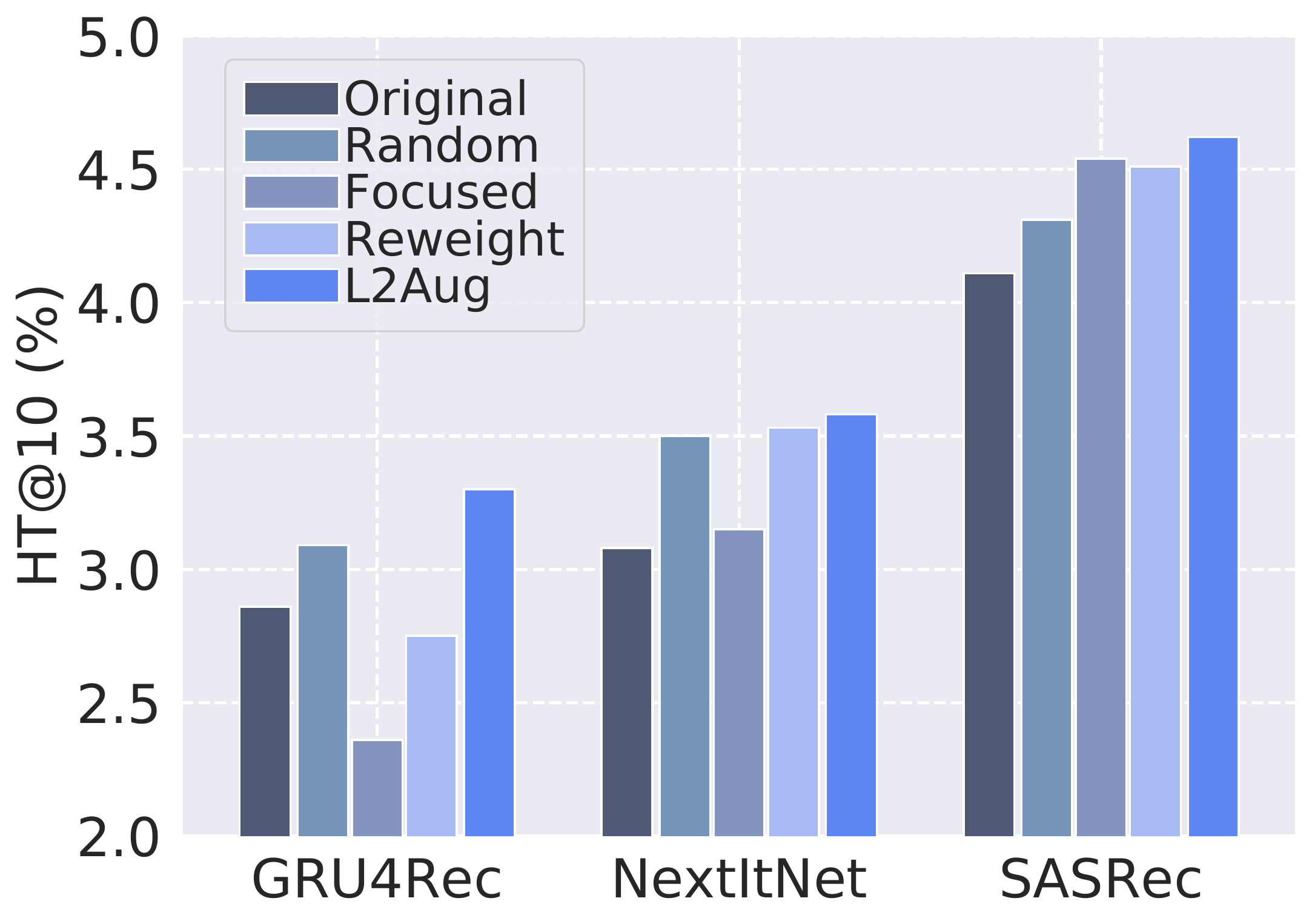}
    }
    \hspace{-0.1cm}
    \subfigure[\textbf{Amazon\_Movies}] 
    {
    \includegraphics[width=0.24\textwidth]{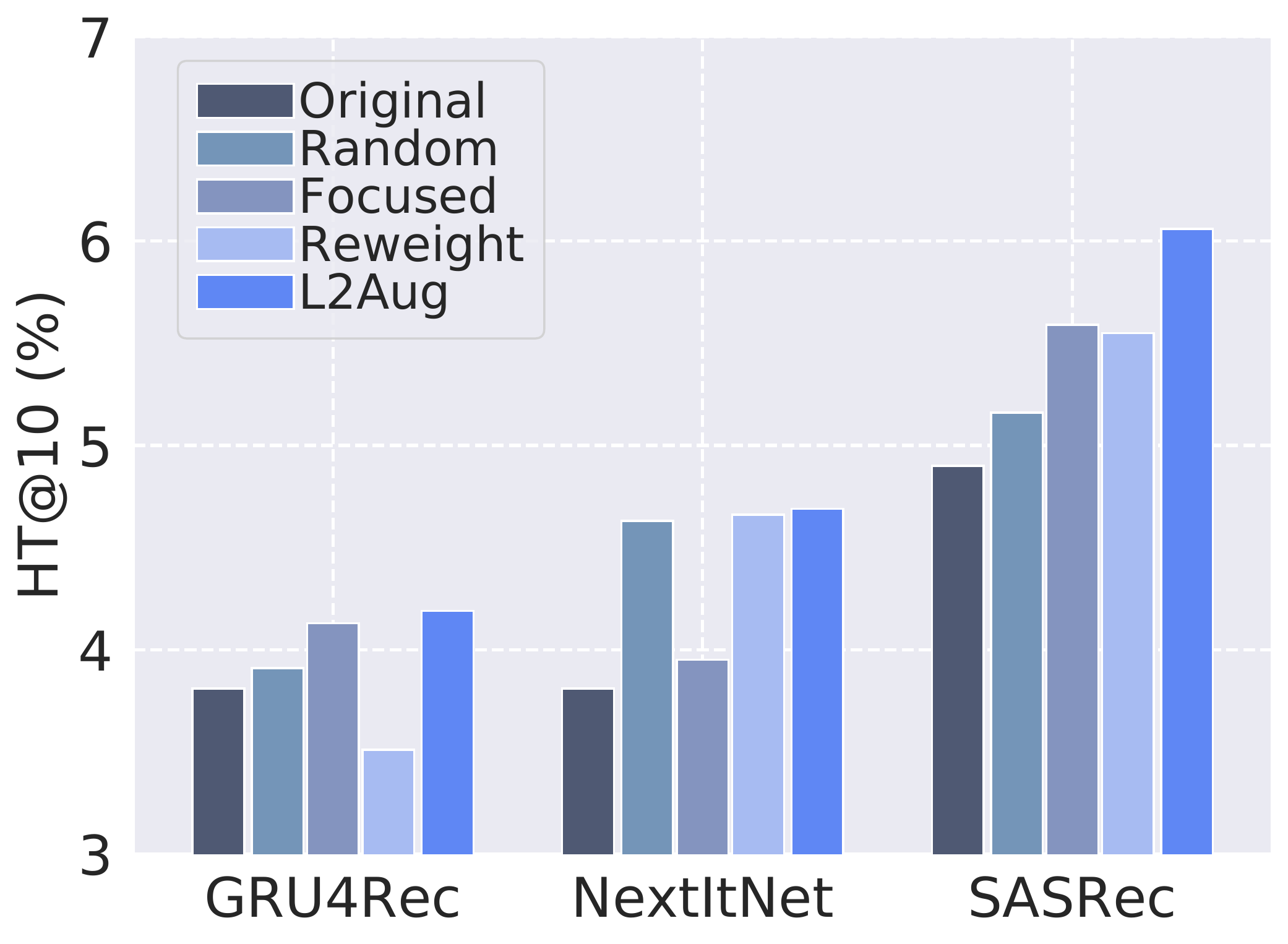}
    }
    \hspace{-0.1cm}
    \subfigure[\textbf{Goodreads}]
    {
    \includegraphics[width=0.24\textwidth]{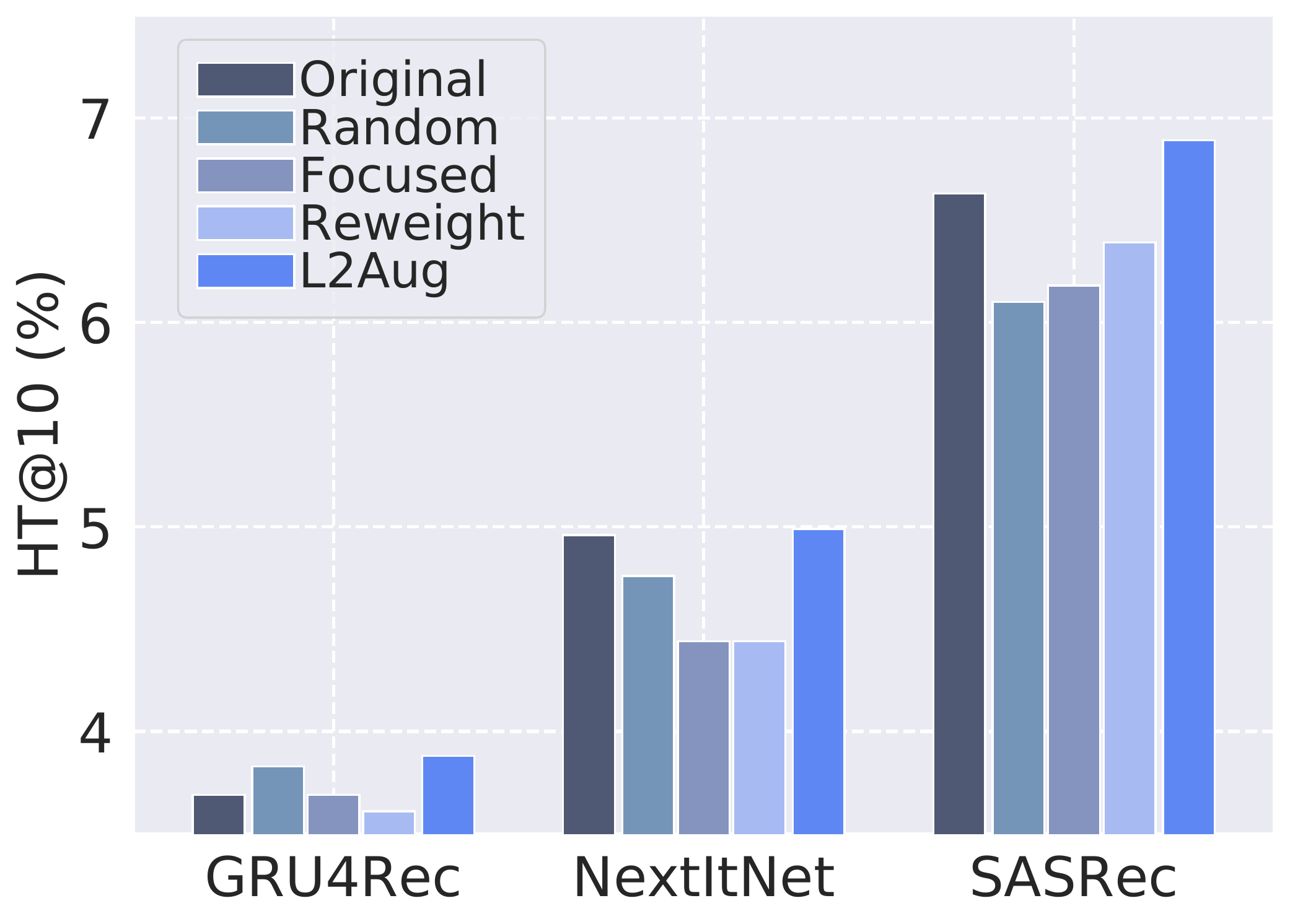}
    }}
     \vspace{-0.1cm}
    \caption{Performance (HT@10) in offline setup on Core User Recommendation of various models on different datasets.}%
    \label{fig:ht10}
\end{figure*}

As supplement to Figure \ref{fig:N}, we compare the performance of \textit{L2Aug} with other treatments for \textit{core users} under HT@5, NDCG@10 and HT@10 in Figure \ref{fig:ht5}, \ref{fig:ndcg10} and \ref{fig:ht10}. Though improving the original sequential recommendation models on casual users, the baseline treatment methods loss their power in modeling core users' transitional patterns and result in a decrease for the overall recommendation. For example, on Amazon\_Books dataset, combining \textit{Focused Learning} with GRURec will result in a $25.05\%$ and $17.48\%$ decrease in core user recommendation. However, the proposed \textit{L2Aug} data augmentation is able to bridge the gap between core and casual users. Thus, we can observe that it can contribute in improvement in both core and casual user recommendation across different datasets.

\end{document}